\documentclass[aps,pra,showpacs,amsmath,amssymb,twocolumn,superscriptaddress]{revtex4-1}
\usepackage{amsmath}
\usepackage{amsfonts}
\usepackage{amssymb}
\usepackage{graphicx}
\graphicspath{{./Figs/}}
\usepackage{epsfig}
\usepackage{color}
\usepackage{empheq}
\usepackage{epstopdf}
\usepackage{braket}
\newcommand{\tmp}[1]{}
\begin{document}
\title{Theory of noiseless phase-mixing amplification in a cavity optomechanical system}

\author{C.~F.~Ockeloen-Korppi}
\affiliation{Department of Applied Physics, Aalto University, P.O. Box 15100, FI-00076 AALTO, Finland}
\author{T.~T.~Heikkil\"a}
\affiliation{Department of Physics and Nanoscience Center, University of Jyvaskyla,
P.O. Box 35 (YFL), FI-40014 University of Jyvaskyla, Finland}
\author{M.~A.~Sillanp\"a\"a}
\affiliation{Department of Applied Physics, Aalto University, P.O. Box 15100, FI-00076 AALTO, Finland}
\author{F.~Massel}
\email[]{francesco.p.massel@jyu.fi}
\affiliation{Department of Physics and Nanoscience Center, University of Jyvaskyla,
P.O. Box 35 (YFL), FI-40014 University of Jyvaskyla, Finland}

% -Limits on the amplification Caves 
%      -phase-sensitive -> can beat noise in one quadrature 
%        (pulsed regime another way) 

% -Why relevant? conceptually-> signal manipulated at a quantum level, detection, QIP (?). 

% -Previous stuff in the microwave regime:
%    - reference to us  (Josephson inductance)
%    - us 
%    - what was done in VTT (?)   
   
% -what we do here

% -why it is good: improves on other schemes dynamic range, bandwith and, of course, noise.

% -nature of the output state -> quantum, statistics.
%       -check Walls & Milburn about this       

\begin{abstract}
  The investigation of the ultimate limits imposed by quantum mechanics on
  amplification represents an important topic both on a fundamental level and
  from the perspective of potential applications. We propose here a novel setup
  for an optomechanical amplifier, constituted by a mechanical resonator
  dispersively coupled to an optomechanical cavity asymmetrically driven around
  both mechanical sidebands. We show that, on general grounds, the present
  amplifier operates in a novel regime-- which we here call phase-mixing
  amplification. At the same time, for a suitable choice of parameters, the
  amplifier proposed here operates as a phase-sensitive amplifier. We show that
  both configurations allow amplification with an added noise below the quantum
  limit of (phase-insensitive) amplification in a parameter range compatible
  with current experiments in microwave circuit optomechanics.
\end{abstract} 

\maketitle 
The amplification of a signal constitutes one of the fundamental technical
aspects of the modality through which modern information and communication
technology operates, potentially paving the way towards the full technological
exploitation of quantum mechanics \cite{Clerk:2010dh}. At the same time it also
represents a fundamental tool in the exploration of the properties of the world
around us: with implications ranging from the exploration of quantum-mechanical
properties of macroscopic objects \cite{ZUREK:1991td} to the detection of
gravitational waves \cite{Abbott:2016ki}. In this context, it is thus relevant
both from a conceptual and the applied point of view to investigate the
boundaries imposed on the amplification of a signal, e.g.  what kind of input
we can effectively amplify and what are the properties of the output given a
specific amplification setup.

In the context of quantum physics, a general result about the limits of
amplification was derived by Haus \cite{Anonymous:rI5t-iRf} and Caves
\cite{Caves:1982wq}, stating that an amplifier, in order for its behaviour to be
consistent with quantum mechanics, must add a minimum amount of noise,
effectively preventing the possibility of cloning a quantum state
\cite{Wootters:1982ex}. In particular, if both quadratures of the input signal
are amplified by the same amount, the minimum added noise corresponds to, in the
large-gain limit, to half a quantum. Below, we refer to this limit as the
amplification quantum limit (AQL) for phase-insensitive amplifiers. In the recent
past, a lot of experimental and theoretical effort has been devoted to the
amplification of quantum signals close to the quantum limit, in particular in
the context of circuit quantum electrodynamics
\cite{CastellanosBeltran:2008cg,Bergeal:2010iu,Zhong:2013ca}, and in
optomechanical setups
\cite{Massel:2011ca,Metelmann:2014bp,Metelmann:2015gb,Toth:2016vk,OckeloenKorppi:2016uy}.

From the theoretical point of view, two possible alternatives have been
contemplated to circumvent this limitation. One relies on the concept of
``nondeterministic noiseless linear amplification'' \cite{Ralph:bx}, according
to which, with a probability of success $p$, it is possible to improve the
signal-to-noise ratio beyond the AQL, with the limiting case of $p=0$ to attain
noiseless amplification. The second idea, dating back to to Haus and Caves'
work, consists in considering a phase-sensitive amplifier, for which, at the
expenses of increased fluctuations in one quadrature, it is possible to reduce
the fluctuations in the other below the AQL imposed on phase-preserving linear
amplifiers.

In this article, we elaborate on the second idea and we report how it is
possbile to reach below-AQL amplification in an optomechanical device suitably
driven by two strong pumping tones. The conceptual relevance of such a device
lies in the fact that it allows a faithful amplification on the level of single
quanta, thus representing an ideal candidate in quantum-information processing
applications, and in the detection of ultraweak signals.  The present amplifier
design possesses other advantages with respect to previous proposals: contrary
to amplifiers based on Josephson junctions (see e.g.
\cite{CastellanosBeltran:2008cg,Bergeal:2010iu,Zhong:2013ca,Lahteenmaki:2014gw})
whose inputs have relatively small dynamic range, the current amplifier works
with comparably large inputs; compared to the optomechanical design proposed in
\cite{Massel:2011ca} the bandwidth is orders of magnitude larger, making this
device, on one hand a pivotal demonstration of how laws of quantum mechanics
shape the properties of amplifiers, and, on the other, a device of unprecedented
power and versatility whose design simplicity make it a perfect candidate for
large-scale technological applications.

Furthermore, we show how the device proposed here can operate in a previously
unreported regime: the analysis of multimode amplifiers has typically focused on
a regime for which each output quadrature solely depends on a specific input
quadrature (on top of the added noise) leading to the definition of
phase-preserving, phase-conjugating and phase- sensitive amplification. Here we
discuss how a more general scenario, in which either output quadrature can depend
on both input quadratures.

\section{Phase-mixing amplification}

The most general example of multimode linear amplifier can be described by the
following input/output relations \cite{Gardiner:1985ig} 
\begin{align}
  a_{{ o} \,\omega  }= &A_{\omega  } a_{{in}\,\omega  } +   B_{\omega  }
                            a^\dagger_{{in}\,-\omega  } +
                            \mathcal{F}_{{in}\, \omega},
  % a^\dagger_{{\rm o} \,-\omega  }= &A^*_{-\omega  } a^\dagger_{{
  %                                     in}\,-\omega  } +  B^*_{-\omega  } a_{{
  %                                    in}\,\omega  } +   \mathcal{F}_{{in}\, -\omega}^\dagger, 
 \label{eq:6r}
\end{align}
where $a_{{\rm in}\,\omega  }$, $a_{{\rm out}\,\omega  }$ and
$\mathcal{F}_{\rm{in}\,\omega}$ represent the operators associated with
the input, output and added noise fields respectively \footnote{The subscript $\omega$
  stands for the frequency, and we use the Fourier convention where
  $a^\dagger_\omega$ is the conjugate of $a_\omega$.}. As discussed in
\cite{Caves:1982wq}, Eq.~\eqref{eq:6r} has to be intended as referred to a
specific carrier frequency $\omega_c$, with respect to which the frequency
$\omega$, and thus the quadratures, are defined (see appendix
\ref{sec:appendix-iii}).  

We can write Eq.~\eqref{eq:6r} in terms
of input $X_\omega^{1,2}$ and output $Y_\omega^{1,2}$ quadratures as
\begin{align}
  Y_\omega^\theta =& \left[A_{11}(\omega) \cos \theta -i A_{21}(\omega) \sin \theta \right] X_\omega^1 + \nonumber \\
                            & \left[i A_{12}(\omega) \cos \theta + A_{22}(\omega) \sin \theta \right]  X_\omega^2  + \mathcal{F}^\theta_\omega,
  \label{eq:7r}
\end{align}
where 
\begin{align}
  &A_{11}\left(\omega\right) = \left[\left(A_\omega+\bar{A}_\omega\right) + \left(B_\omega+\bar{B}_\omega\right)\right]/2 \nonumber \\ 
  &A_{12}\left(\omega\right) = \left[\left(A_\omega-\bar{A}_\omega\right) - \left(B_\omega-\bar{B}_\omega\right)\right]/2, \nonumber \\
  &A_{21}\left(\omega\right) = \left[\left(A_\omega-\bar{A}_\omega\right) + \left(B_\omega-\bar{B}_\omega\right)\right]/2 \nonumber \\
  &A_{22}\left(\omega\right) = \left[\left(A_\omega+\bar{A}_\omega\right) -\left( B_\omega+\bar{B}_\omega\right)\right]/2  
 \label{eq:8r}
\end{align}
with $X_\omega^1=a^\dagger_{{\rm in}\,-\omega}+a_{{\rm in}\,\omega}$,
$X_\omega2=i \left(a^\dagger_{{\rm in}\,-\omega}-a_{{\rm in}\,\omega}\right)$,
$Y_\omega^\theta=\left(a^\dagger_{{\rm in}\,-\omega}e^{i\theta}+a_{{\rm
      in}\,\omega}e^{-i\theta}\right)$,
$\mathcal{F}^\theta_\omega=\left(\mathcal{F}^\dagger_{{\rm
      in}\,-\omega}e^{i\theta}+\mathcal{F}_{{\rm
      in}\,\omega}e^{-i\theta}\right)$, and
$\bar{A}_\omega,\bar{B}_\omega=A^*_{-\omega},\,B^*_{-\omega} $. The phase
$\theta$ represents a controllable parameter, related to the homodyne detection
scheme characterising phase-sensitive measurements both in the optical and in
the microwave regime \cite{Walls:2008em}.

Defining $Y_1=Y_\omega^{\pi/2}$ and $Y_2=Y_\omega^{0}$, we can write Eq. \eqref{eq:7r} in matrix form 
\begin{align}
  \mathbf{Y}= \mathbf{A} \mathbf{X} +\mathbf{\mathcal{F}}
  \label{eq:9r}  
\end{align}
with  $\mathbf{A}=[A_{11},i A_{12};-i A_{21},A_{22} ]$, $\mathbf{Y}=\left[Y_1 , Y_2\right]^T$, $\mathbf{X}=\left[X_1 , X_2\right]^T$,
$\mathbf{\mathcal{F}}=\left[\mathcal{F}_1 , \mathcal{F}_2\right]^T$.

Equation \eqref{eq:9r} constitutes a generalisation of the analysis performed by
Caves in the sense that we do not constrain the coefficients of
Eq. \eqref{eq:6r} (and the corresponding equation for
$a^\dagger_{\rm{o}\,-\omega}$) to obey the relation $A^*_{-\omega}=A_{\omega}$,
$B^*_{-\omega}=B_\omega$, as in the case discussed by Caves for multimode
phase-sensitive amplifiers (see discussion before Eqs. 4.40 in
ref. \cite{Caves:1982wq}), for which $A_{12}$ and $A_{21}$ would be identically
zero. 

In order to characterise the deviation from the case of multimode
phase-sensitive amplification, we write the coefficients $A_\omega$ and
$B_\omega$ in terms of their symmetric and antisymmetric frequency components
\begin{align}
  A_\omega = A_{\Sigma\, \omega} + A_{\Delta\, \omega} \nonumber \\
  B_\omega = B_{\Sigma\, \omega} + B_{\Delta\, \omega},
  \label{eq:1n}
\end{align}
where $A_{\Sigma\, \omega}=\left(A_\omega+ A_{-\omega}\right)/2$,
$A_{\Delta\, \omega}=\left(A_\omega-A_{-\omega}\right)/2$ and analogously for
$B$. In addition we exploit the gauge freedom for the input
($a_{\mathrm{i} \omega} \to a_{\mathrm{i} \omega} \exp \left[i \phi_{{\rm i}
    \omega}\right]$) and output fields
($a_{\mathrm{o} \omega} \to a_{\mathrm{o} \omega} \exp \left[i \phi_{{\rm o}
    \omega}\right]$) imposing that
\begin{align}
  \phi_{\Sigma\,\omega}^A&=\phi_{{\rm o} \omega}-\phi_{{\rm i} \omega} \nonumber \\
  \phi_{\Sigma\,\omega}^B&=\phi_{{\rm o} \omega}-\phi_{{\rm i} -\omega} = \phi_{{\rm o} \omega}+\phi_{{\rm i} \omega}
  \label{eq:2n}
\end{align}
where $\phi_{\Sigma\, \omega}^A={\rm Arg}\left[A_{\Sigma\,\omega}\right]$, 
$\phi_{\Sigma\, \omega}^B={\rm Arg}\left[B_{\Sigma\,\omega}\right]$. 
We can write the equations of motion in a rotated frame, for which
\begin{align}
   \mathbf{Y}=\mathbf{\tilde{A}} \mathbf{X}
    \label{eq:3n}
\end{align}
where 
\begin{align}
   \tilde{A}_{11} &= \left|A_{\Sigma\, \omega}\right| +
                         \left|B_{\Sigma\, \omega}\right| +
                       i \left(
                                \left|A_{\Delta\,\omega}\right| \cos \phi_1
                             +\left|B_{\Delta\, \omega}\right| \cos \phi_2 
                         \right) \nonumber \\
  \tilde{A}_{12} &= i \left(
                                \left|B_{\Delta\,\omega}\right| \cos \phi_2
                             - \left|A_{\Delta\, \omega}\right| \cos \phi_1 
                         \right) \nonumber \\
\tilde{A}_{21} &=   i \left(
                                \left|A_{\Delta\,\omega}\right| \cos \phi_1
                             +\left|B_{\Delta\, \omega}\right| \cos \phi_2 
                         \right) \nonumber \\
\tilde{A}_{22} &= \left|A_{\Sigma\, \omega}\right| -
                         \left|B_{\Sigma\, \omega}\right| +
                       i \left(
                                \left|A_{\Delta\,\omega}\right| \sin \phi_1
                             -\left|B_{\Delta\, \omega}\right| \sin \phi_2 
                         \right) 
\label{eq:4n}
\end{align}
with
$\phi_1={\rm Arg}\left[A_{\Delta\,\omega}\right]-{\rm
  Arg}\left[A_{\Sigma\,\omega}\right]$ and
$\phi_2={\rm Arg}\left[B_{\Delta\,\omega}\right]-{\rm
  Arg}\left[B_{\Sigma\,\omega}\right]$.  We note that, if
$A_{\Delta\,\omega}=B_{\Delta\,\omega}=0$, Eq. \eqref{eq:4n} corresponds to the
usual input/output relation for a phase sensitive amplifier in the preferred
quadratures.

In terms of algebraic properties, the possibility of diagonalising the matrix
$\mathbf{A}$ through a phase rotation of the input and output fields,
corresponding to a rotation of the quadratures $\mathbf{X} \to R_X \mathbf{X}$,
$\mathbf{Y} \to R_Y \mathbf{Y}$ is equivalent to the statement that for each
real matrix $\mathbf{M}$ there exists the singular value decomposition (SVD)
\begin{align}
  \mathbf{M}= \mathbf{U} \mathbf{D} \mathbf{V}^\dagger
  \label{eq:5n}
\end{align}
where $\mathbf{D}$ is a diagonal matrix and $\mathbf{U}$ and $\mathbf{V}$ are
orthogonal matrices. However, if $\mathbf{M}$ is a complex matrix, the SVD is
possible only in terms of unitary matrices. Since a unitary transformation does
not necessarily map quadrature operators to quadrature operators --the most
prominent example being the mapping between $a$, $a^\dagger$ and (normalised)
quadrature operators-- we are led to conclude that, in general, not all matrices
describing linear amplifiers can be put in a preferred quadrature form: more
specifically due to the residual gauge freedom in the definition of input and
output phases, the only transformations allowed are those defined by orthogonal
matrices modulo an overall complex phase factor. We designate the regime 
for which it is not possible to cast the input-output relations for the field
quadratures as phase-mixing amplification (PMA).

In addition, from the expression of the matrix elements given in
Eq. \eqref{eq:4n}, we note that $\mathbf{\tilde{A}}$ is a diagonal matrix for
$\omega=0$ and thus we recover the usual input/output expressions for a
narrowband phase-sensitive linear amplifier \tmp{maybe expand}
\begin{align}
  Y_1 = A_{11} X_1+ \mathcal{F}_1 \nonumber \\
  Y_2 = A_{22} X_2 + \mathcal{F}_2. 
  \label{eq:19r}
\end{align}
This shows that PMA devices are intrinsically multimode amplifiers.

While we elaborate more about the noise analysis in the specific case of the
optomechanical PMA, we note here that its analysis is somewhat complicated by
the fact that the output in each quadrature depends on both input quadratures.
In general, we can write the output power spectrum as
\begin{align} 
     S_Y^\theta = &O^\theta_1  S_{1} + O^\theta_2 S_{2} + S^\theta_F 
  \label{eq:2u}
\end{align}
where \tmp{check whether the factors 1/2 vs 1/4 are ok in the optomech case}
$S_Y^{\theta}=\frac{1}{4} \left(\braket{Y^\theta_{-\omega}
    Y^\theta_\omega}-1\right)$, and analogously for the input and noise
spectra. From Eqs. (\ref{eq:8r}, \ref{eq:9r}), we obtain
\begin{align}
  O^\theta_1 =& \left[\left|A_{11}\right|^2    \cos^2 \theta +
                             \left|A_{21} \right|^2   \sin^2 \theta\right]
  \nonumber \\
&+ \sin 2\theta  \left|A_{11}A_{21}\right| \sin \left[\phi_{21}-\phi_{11}\right]
  \nonumber\\
  O^\theta_2 =& \left[\left|A_{22}\right|^2    \sin^2 \theta +
                             \left|A_{12} \right|^2  \cos^2 \theta \right] \nonumber \\
                     &- \sin 2\theta  \left|A_{22}A_{11}\right| \sin \left[\phi_{12}-\phi_{22}\right] 
  \label{eq:3u}
\end{align}
with $\phi_{\rm ij}={\rm Arg}\left[A_{\rm ij}\right]$.  In order to simplify our
analysis, we will consider here an input for which $S_{1}=S_{2}$ (i.e. we
exclude from our noise analysis the possibility of a squeezed input state) and
therefore define the PMA power gain as
\begin{align}
  \left|\mathcal{G}^\theta\right|^2= O^\theta_1 + O^\theta_2. 
  \label{eq:4u}
 \end{align}
This allows us to evaluate the added noise as referred to the input as 
\begin{align}
   S_X^\theta = \left. \frac{S_Y^\theta}{\left|\mathcal{G}^\theta\right|^2} \right|_{S_1,S_2=0}
                 = \left. \frac{S_Y^\theta}{O^\theta_1 + O^\theta_2} \right|_{S_1,S_2=0}.
  \label{eq:5u}
\end{align}

\section{Optomechanical PMA}
\label{sec:optomechanical-pma}

The setup considered to demonstrate PMA is represented by the prototypical, and
arguably most simple, optomechanical cavity system, consisting of an
electromagnetic (optical or microwave) cavity with resonant frequency $\omega_c$
dispersively coupled to a mechanical oscillator whose resonance frequency is
given by $\omega_m$ (see e.g. \cite{Aspelmeyer:2014ce}). 
\begin{figure}[ht]
\includegraphics[width=0.5\textwidth]{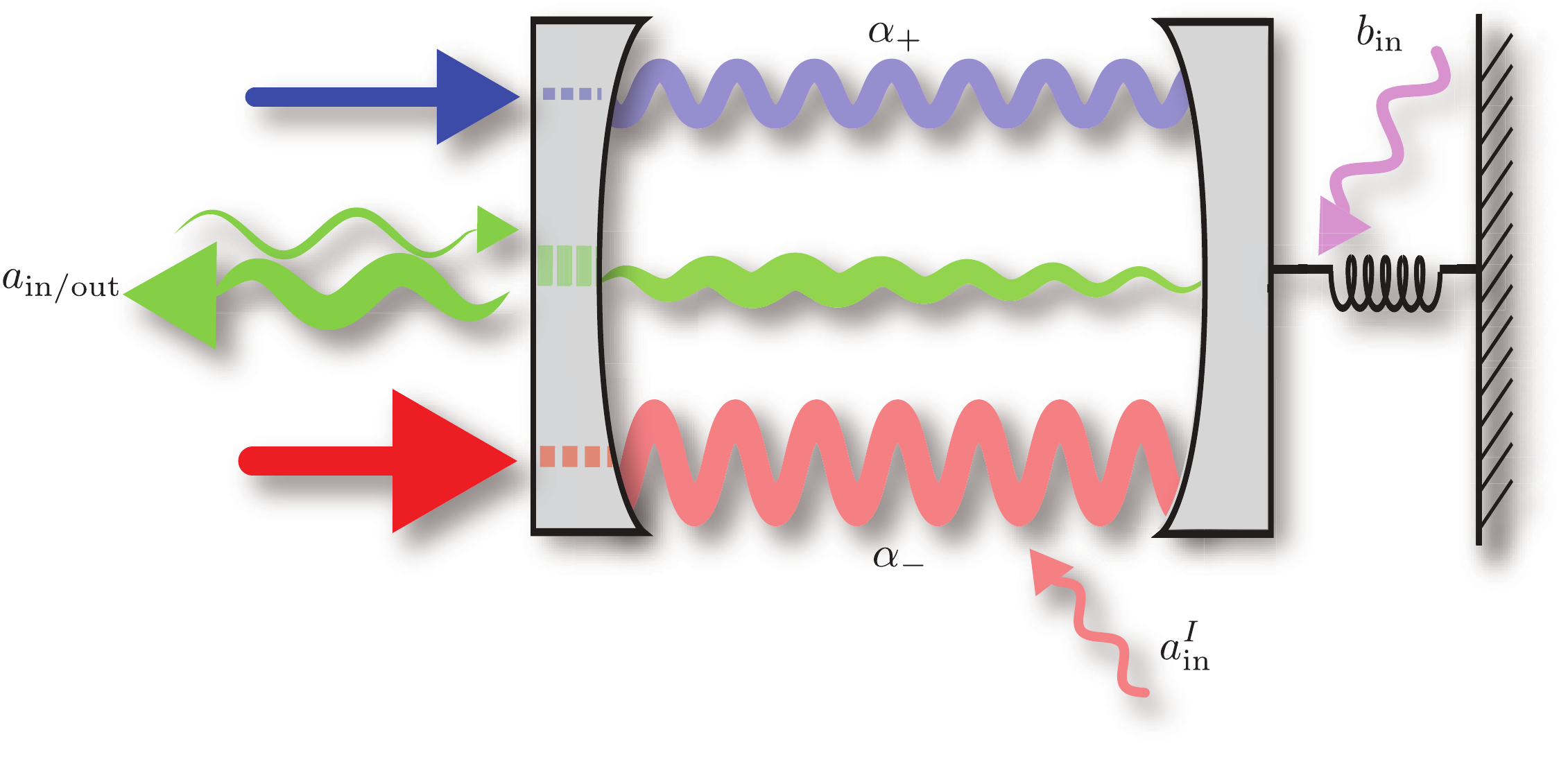}
\caption{Schematic representation of the setup discussed here. The cavity is driven
  by two tones at frequencies and intensities $\omega_+$, $\alpha_+$ (blue in
  the figure) and $\omega_-$, $\alpha_-$, respectively (in red). In the figure we
  also indicate the input and output signals ($a_{\rm in}$, $a_{\rm out}$), and the internal
and mechanical noise ($a^I_{\rm in}$ and $b_{\rm in}$).}
\label{cavity}
\end{figure}

The Hamiltonian of the system can be written as
\begin{align} 
H= \omega_c a^\dagger a + \omega_m b^\dagger b + g_0 a^\dagger a
\left( b^\dagger +b \right),
  \label{eq:1r}
\end{align}
where $a$ ($a^\dagger$) and $b$ ($b^\dagger$) represent the raising (lowering)
operators associated with the electromagnetic cavity field and the mechanical
oscillator, respectively, and $g_0$ is the single-photon optomechanical coupling
strength.  In addition to its internal dynamics, the system is coupled to an
environment, which provides the possibility of driving and probing the system
and, at the same time, represents a source for noise and dissipation, both for
the mechanics and the cavity.  Furthermore, we describe the noise/dissipation
properties of the mechanical resonator through the coupling with a (phononic)
thermal reservoir with average population $n_m$, and define a characteristic
linewidth $\gamma$. An analogous assumption is adopted for the cavity. In this
case, however, we consider a coupling to two different baths: the external bath
(characterized by the linewidth $\kappa_e$) providing both input signal and
input noise, and an internal bath (linewidth $\kappa_i$), associated with the
internal losses of the cavity and whose population is given by $n_c^I$.
\begin{figure}[ht]
\includegraphics[width=0.5\textwidth]{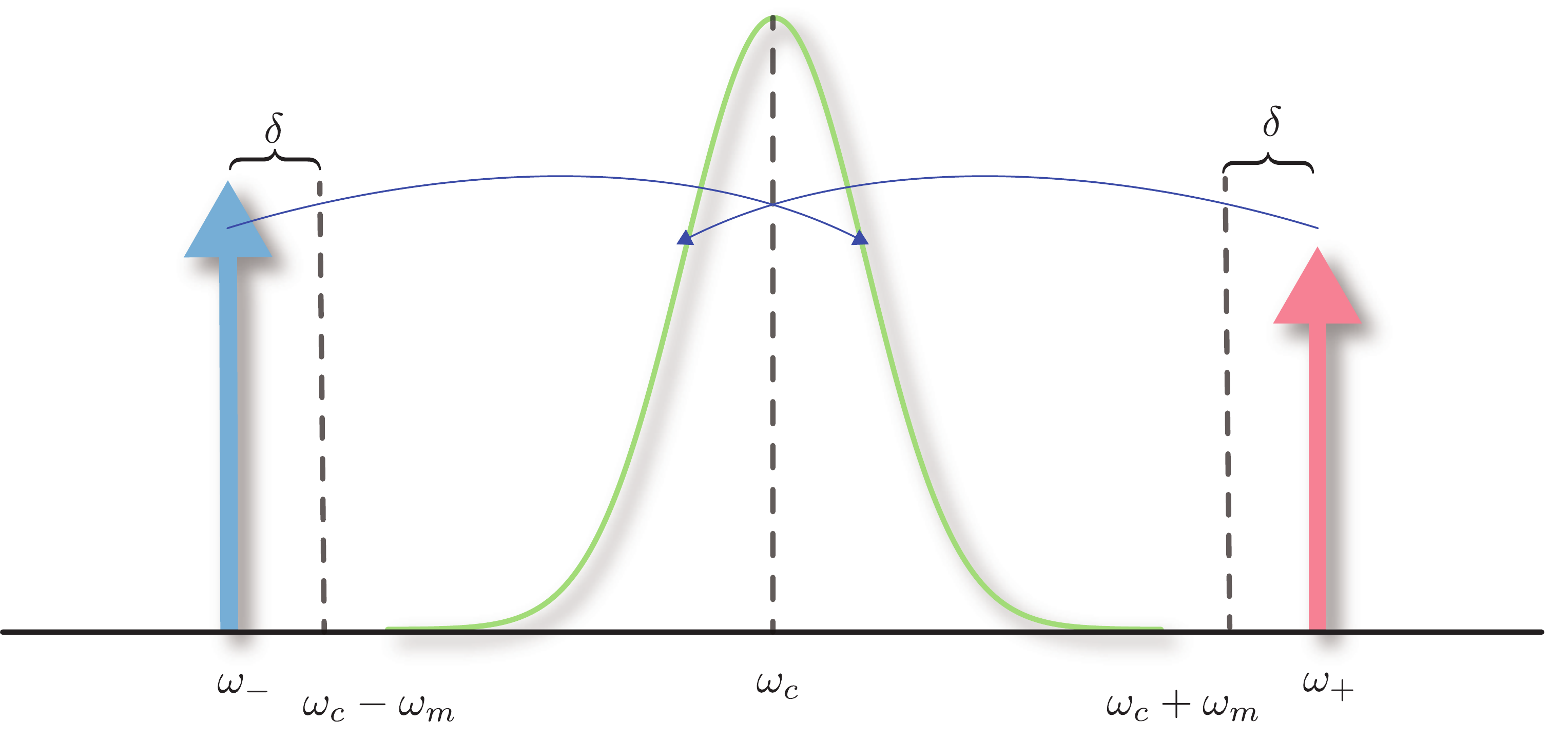}
\caption{Schematic representation of the pump intensities and frequencies with
  respect to the cavity frequency and linewidth. The mechanics-mediated
  scattering of pump photons results in the amplification of electromagnetic
  signals scattered from the cavity around the cavity frequency resonance
  $\omega_c$.}
\label{freqs}
\end{figure}
Concerning the driving of the system, we assume that the cavity is driven by two
strong pumps of amplitude $\alpha_+$ and $\alpha_-$, which are detuned with
respect to the cavity resonant frequency $\omega_{\rm c}$, by
$\omega_+ -\omega_c= \omega_{\rm m} + \delta$ and
$\omega_- -\omega_c= -\omega_{\rm m} -\delta$, respectively (Fig. \ref{freqs}).
A related two-tone setup has been previously considered in the context of backaction
evading (BAE) measurements of the mechanical oscillator position
\cite{Caves:1980jpa,Braginsky:1980hj,Clerk:2008je,Hertzberg:hf}, and in the
generation of mechanical squeezing \cite{Wollman:2015gx, Pirkkalainen:2015ki,
  Lecocq:2015dk}. In both cases the frequencies of the driving tones were
considered to fulfil the relation $\omega_\pm=\omega_c \pm \omega_m$. For equal
pump amplitudes ($\alpha_+=\alpha_-$) this leads to the BAE detection of the
mechanical oscillator position, and for $\alpha_+\neq \alpha_-$ to the squeezing
of the mechanics below the standard quantum limit, defined as the uncertainty
associated with the ground state of the mechanical oscillator.

In the presence of two strong driving tones, we can follow a standard approach
and linearise the Hamiltonian given in Eq. \eqref{eq:1r}.
% \begin{align}
%   H=&\omega_{c} a^\dagger a + \omega_{c} b^\dagger b  \nonumber\\ 
%                     +&\left(G_+ e^{-i\omega_+ t} +G_- e^{-i\omega_- t} \right) \left(a^\dagger b^\dagger + a^\dagger
%                       b\right) +  \mathtt{h.c.},
% \label{eq:2}                      
% \end{align}
Neglecting fast oscillating terms (rotating-wave approximation) and
moving to a frame rotating at $\omega_{c}$ and $\omega_{m}-\delta$ for the
cavity and the mechanical field, respectively, we can write it as
\begin{align}
  H= \delta\, b^\dagger b + G_+ a^\dagger b^\dagger +G_- a^\dagger b +  {h.c.}.
  \label{eq:2r}
\end{align}
where $G_\pm= g_0 \alpha_\pm$.

The solution of the equations of motin becomes simple after expressing
Eq. \eqref{eq:2r} in terms of Bogoliubov modes for the cavity field
\begin{align}
  H= \delta\, b^\dagger b 
   +  G_{{BG}}\left(\alpha^\dagger b + \alpha b^\dagger \right),
  \label{eq:3r}
\end{align}
where $\alpha=u \, a + v \, a^\dagger$,
$G_{BG}=(G_-^2-G_+^2)^{1/2}$, $u=G_-/G_{BG}$, $ v =G_+/G_{BG}$.
The beam-splitter term
$ G_{{BG}}\left(\alpha^\dagger b + \alpha b^\dagger \right)$ in
Eq. \eqref{eq:3r} points towards the cooling of the mechanical motion to the
temperature of the Bogoliubov cavity mode. As we show below, this entails the
amplification of the unrotated cavity mode $a$.

From Eq. \eqref{eq:3r} we can determine the following quantum Langevin
equations in the frequency domain for $\alpha$
and $b$
\begin{align}
    -i \omega \alpha_\omega &= - i G_{{BG}} b_\omega
  -\frac{\kappa}{2} \alpha_\omega + \sqrt{\kappa_e} \alpha_{{in}\,\omega} + \sqrt{\kappa_i} \alpha^I_{{in}\,\omega}\nonumber \\
  -i \omega b_\omega &= -i \delta b_\omega + i G_{{BG}} \alpha_\omega - \frac{\gamma}{2} b_\omega + \sqrt{\gamma} b_{{in}\,\omega}.
\label{eq:4r}
\end{align}
Eliminating the mechanical degrees of freedom from Eq. \eqref{eq:4r},
considering the usual input-output relation
$\alpha_{{\rm o}\,\omega}+\alpha_{{\rm in}\,\omega}=\sqrt{\kappa}
\alpha_\omega$, and transforming back to $a_\omega$, we can obtain an
input/output relation for the output field $a_{{\rm o} \,\omega}$(see Appendix
\ref{sec:appendix-i})
\begin{align}
    a_{{\rm o} \,\omega  }= &A_{\omega  } a_{{\rm in}\,\omega  } + 
                                                 B_{\omega  } a^\dagger_{{\rm in}\,-\omega  } +  
                                                 A_{I\,-\omega } a^{I}_{{\rm in}\,\omega  }  +  
                                                 B_{I\,\omega }  {a^{I}}^\dagger_{{\rm in}\,-\omega  } +   \nonumber  \\
                                              &C_{\omega  } b_{{\rm in}\,\omega  } + D_{\omega  } b^\dagger_{{\rm in}\,-\omega  }. 
   % a^\dagger_{{\rm o} \,-\omega  }= &A^*_{-\omega  } a^\dagger_{{\rm in}\,-\omega  } + 
   %                                               B^*_{-\omega  } a_{{\rm in}\,\omega  } +  
   %                                               A^*_{I\,-\omega } a^{I}_{{\rm in}\,\omega  }  +  
   %                                               B^*_{I\,-\omega }  a^{I}_{{\rm in}\,\omega  } +   \nonumber  \\
   %                                            &C^*_{-\omega  } b^\dagger_{{\rm in}\,-\omega  } + D^*_{-\omega  } b_{{\rm in}\,\omega  },
  \label{eq:5r}
\end{align}
The coefficients in Eq. \eqref{eq:5r} 
are given by
\begin{align}
  &A_\omega= \kappa_{\rm e}\left( u^2  \chi_{\rm c}^{\rm eff} - v^2 \bar{\chi}_{\rm c}^{\rm eff}\right) -1 \nonumber \\
  &A_{I\,\omega}=   \sqrt{\kappa_{\rm i} \kappa_{\rm e}}\left( u^2  \chi_{\rm c}^{\rm eff} - v^2 \bar{\chi}_{\rm c}^{\rm eff}\right) \nonumber \\
  &B_\omega=u v  \kappa_{\rm e} \left(\chi_{\rm c}^{\rm eff} - \bar{\chi}_{\rm c}^{\rm eff}\right)  \nonumber \\
  &B_{I\,\omega}= u v \sqrt{\kappa_{\rm i} \kappa_{\rm e}}  \left(\chi_{\rm c}^{\rm eff} - \bar{\chi}_{\rm c}^{\rm eff}\right)  \nonumber \\
  &C_\omega=-i G_- \sqrt{\gamma \kappa_{\rm e}}  \chi_{\rm c}^{\rm eff} \chi_m  \nonumber  \\
  &D_\omega=i G_+ \sqrt{\gamma \kappa_{\rm e}}  \bar{\chi_{\rm c}^{\rm eff}} \bar{\chi}_m
  \label{eq:10r}
\end{align}
with $ \bar{\chi} = \chi^*(\omega \to -\omega) $,
$\chi=\chi_c^{\mathrm{eff}},\chi_m $, and
\begin{align}
 % & \sqrt{\tilde{\kappa}} \tilde{\alpha}_{{in}\,\omega}=\sqrt{\kappa} \alpha_{{in}\,\omega}-i G_{{BG}}
%\chi_m \sqrt{\gamma}b_{{in}\,\omega} \nonumber \\
&\chi_c^{\mathrm{eff}} =\left[\kappa/2-i\omega+G_{{BG}}^2\chi_m\right]^{-1},
  \nonumber \\
&\chi_m=\left[\gamma/2 -i \left(\omega-\delta\right) \right]^{-1}.
\label{eq:11r}
\end{align}

\begin{figure}[ht]
\includegraphics[width=0.5\textwidth]{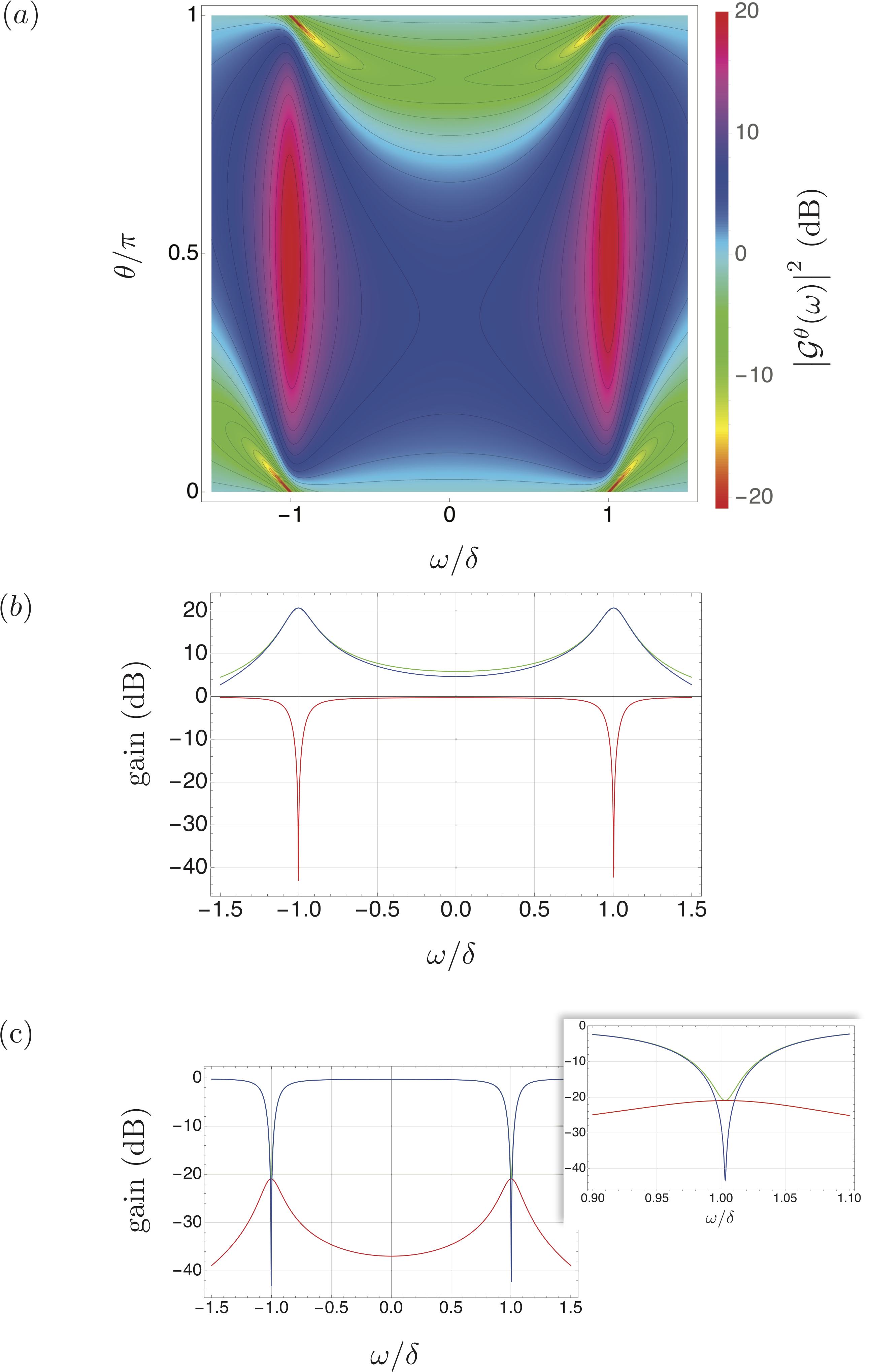}
\caption{\emph{(a)} Gain for below-AQL amplification (see Eq. \eqref{eq:17r} below) as
   a function of $\theta$ and $\omega$.  Detail of the frequency
  dependence around $\omega_{max}$ of $\left|\mathcal{G}^\theta(\omega)\right|^2$ (green),
  $O^\theta_1(\omega)$ (blue), $O^\theta_2(\omega)$ (red), for
  $\theta=\pi/2$\emph{(b)} and $\theta=0$\emph{(c)}. Parameters:
  $G_+= 0.06$, $ G_-= 0.072$, $\delta=0.04$,
  $\kappa_{e}=0.99$, $\gamma=2 \cdot 10^{-5}$, energies in units of the
  cavity linewidth $\kappa$. The frequency $\omega=0$ corresponds to the cavity
  resonant frequency.}
\label{GainAbove3D}
\end{figure}

Equations (\ref{eq:10r},\ref{eq:11r}) allow us to identify, for the
optomechanical case, the parameters defined in Eq. \eqref{eq:9r}. More
specifically, the definitions given in \eqref{eq:10r} allow us to evaluate
$O_1^\theta$, $O_2^\theta$, and $S_F$, therefore characterising the PMA
properties of the system. \tmp{Elaborate here} In Fig. \ref{GainAbove3D} we
characterise the phase-mixing properties of the amplifier. In particular it is
possible to see that, at the maximum-gain frequency $\omega_{max}$ (see
Eq. \eqref{eq:13r} below) we have that
$\left|\mathcal{G}^{\pi/2}(\omega)\right|^2 \simeq O^{\pi/2}_1$
(Fig. \ref{GainAbove3D}(b)) and
$\left|\mathcal{G}^0(\omega)\right|^2 \simeq O^0_2$ (Fig. \ref{GainAbove3D}(c)).
From Eqs. (\ref{eq:3u},\ref{eq:4u}), this implies the coefficients $A_{11}$ and
$A_{22}$ are negligible with respect to the diagonal terms, and therefore allow
us to describe the device as a phase sensitive amplifier,
\begin{align}
  Y_1 \simeq A_{12} X_2 + \mathcal{F}_2 \nonumber \\
  Y_2 \simeq A_{21} X_1 + \mathcal{F}_1.
  \label{eq:6u}  
\end{align}
% Since it is possible to freely rotate the input and the output quadratures, for
% sake of clarity we will rotate here the output quadratures defining
% \begin{align}
%   Y_1 \simeq \mathcal{G}_1 X_1 + \mathcal{F}_1 \nonumber \\
%   Y_2 \simeq -\mathcal{G}_2 X_2 + \mathcal{F}_2.
%   \label{eq:6bu}   
% \end{align}

% with
% \begin{align}
%   \mathcal{G}_1 =A_{21} \nonumber \\
%   \mathcal{G}_2 =A_{12}.
%   \label{eq:7u}  
% \end{align}
If we are in a sideband resolved-like regime, i.e. if the two peaks depicted in
Fig. \ref{GainAbove3D} can be
approximately treated as separate peaks for
$\kappa_{e} \simeq \kappa$ and $\gamma \simeq 0$, it is possible to
express the gain in terms of a Lorentzian centered around $\omega_{max}$ and
linewidth $\gamma_{eff}$, where
\begin{align}
 \omega_{max}&=\pm \delta \left[1+ \frac{G_{BG}^2}{\kappa^2/4+\delta^2}\right]
  \nonumber \\
  \gamma_{eff}& = \frac{G_{BG}^2\kappa}{\kappa^2/4+\delta^2}.
 \label{eq:13r}
 \end{align}
 These expressions are hence valid for $\omega_{max} \gg \gamma_{eff}$.
 Crucially, for the description of this optomechanical system in terms of PMA,
 away from the resonance defined by Eq. \eqref{eq:13r} the mixing coefficients
 $A_{11}$ and $A_{22}$ start to play a significant role (see
 Fig. \ref{GainAbove3D}), and a real-valued SVD decomposition becomes, in
 general, not possible.
In the limit $G_{BG} \ll \delta$, $\delta\ll \kappa$
the coefficients $A_{ij}$ assume a particularly simple form 
\begin{align}
A_{21} &= -\frac{2/\kappa  \left(G_-+G_+\right)^2}{\frac{\gamma_{eff}}{2} - i
  \left(\omega-\omega_{max}\right)}\nonumber \\
A_{12} &=-\frac{2/\kappa \left(G_--G_+\right)^2}{\frac{\gamma_{eff}}{2} - i
  \left(\omega-\omega_{max}\right)}\nonumber \\
A_{11}&=A_{22}=\left[1-\frac{2 G_{BG}^2/\kappa}{\frac{\gamma_{eff}}{2} - i
  \left(\omega-\omega_{max}\right)}\right].
\label{eq:12r} 
\end{align}
Eqs. (\ref{eq:13r},\ref{eq:12r}) allow us to evaluate an approximate expression
for the gains at resonance ($\omega=\omega_{max}$)
\begin{align}
 \left|\mathcal{G}_1\right|= \left( u +  v \right)^2\nonumber \\
 \left|\mathcal{G}_2\right|= \left( u - v \right)^2 
  \label{eq:16r}
\end{align}
and therefore the value of the gain-bandwidth product
\begin{align}
   \left. \mathcal{G}_1\gamma_{eff} \right|_{\omega = \omega_{max}} = 16 \frac{G_+^2 G_-^2}{\kappa
  G_{BG}^2}.
  \label{eq:8u}
\end{align}
Furthermore,  given the definitions of $u$ and $v$, which can be also expressed as
$u=\cosh \xi$, $v=\sinh \xi$, we can recover the condition 
\begin{align}
  \left| \mathcal{G}_1 \mathcal{G}_2 \right|=1
  \label{eq:9u}
\end{align}
characterising a degenerate parametric amplifier, which can be considered as the
``gold standard'' of phase-sensitive amplifiers.  Furthermore, it is clear from
Eq. \eqref{eq:12r} that, in the limit discussed here, the frequency range around
$\omega_{max}$ for which the system can be characterised as a phase-sensitive
amplifier is given by $\gamma_{eff}$. \tmp{doesn't totally look like that
  from Fig. \ref{GainAbove3D}}

\begin{figure}[ht]
\includegraphics[width=0.4\textwidth]{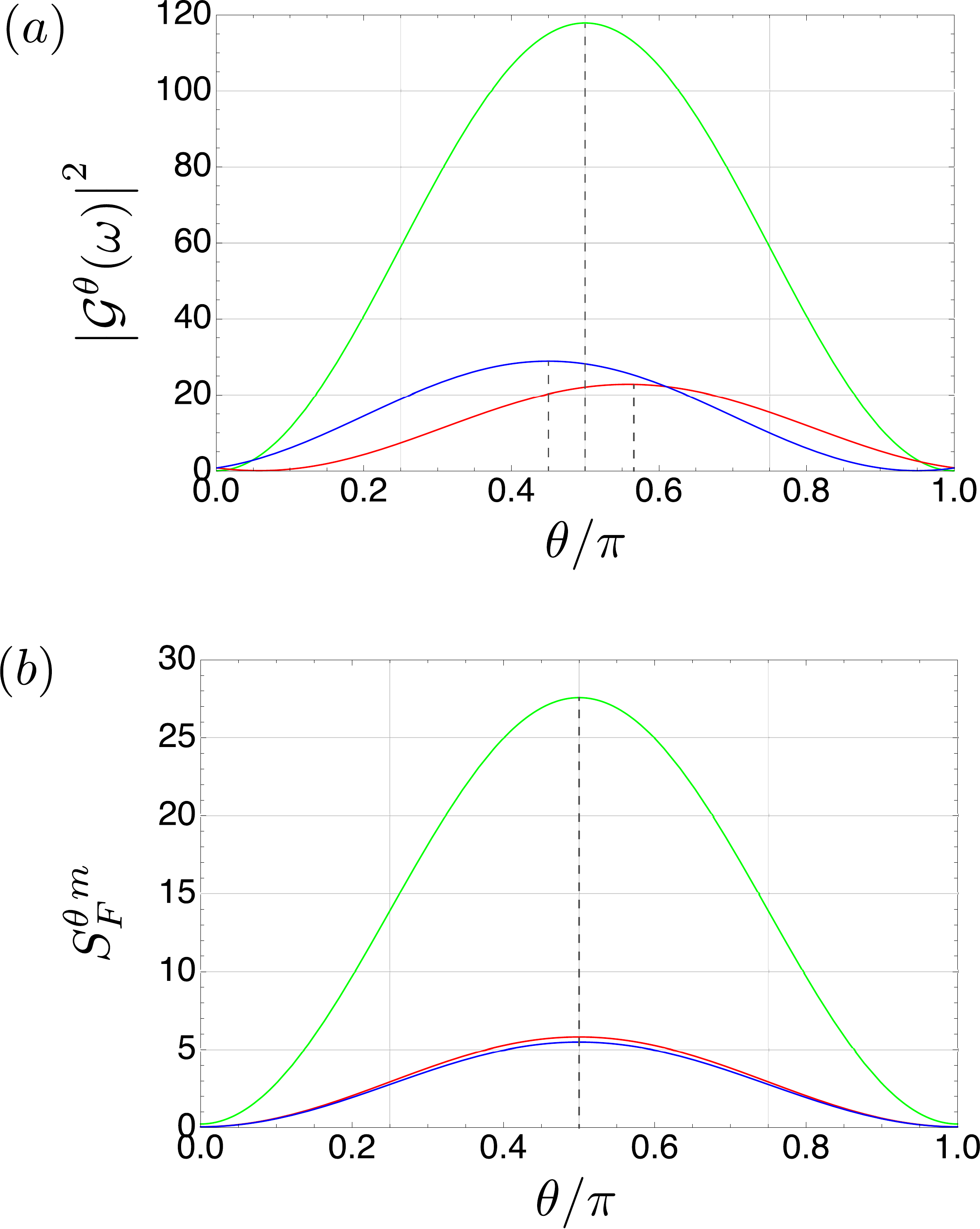}
\caption{\emph{(a)} Gain $\left|\mathcal{G}^\theta(\omega)\right|^2$ for the same
  parameters as in Fig. \ref{GainAbove3D}, for $\omega=\omega_{max}$ (green),
  $\omega=\omega_{max}- \gamma_{eff}$ (red), $\omega=\omega_{max}+ \gamma_{eff}$
  (blue). The maximum point is at $\theta=\pi/2$ for $\omega=\omega_{max}$, and
  shifted from this point for $\omega=\omega_{max} \pm \gamma_{eff}$. \emph{(b)}
  Mechanical contribution to the added noise $S^{\theta\,m}_F$ for
  $\omega=\omega_{max}$ (green), $\omega=\omega_{max}- \gamma_{eff}$ (red),
  $\omega=\omega_{max}+ \gamma_{eff}$ (blue). In this case, the maximum point
  lies at $\theta$ for all values of $\omega$.}
\label{GainAbove} 
\end{figure}
\tmp{2 $\gamma_{eff}$, change caption} In Fig. \ref{GainAbove} we plot the gain
$\left|\mathcal{G}^\theta(\omega)\right|^2$ as a function of $\theta$ for
different values of $\omega$. The crucial feature of this plot is the
$\omega$-dependence of the gain maximum. This dependence, which plays an
important role in the determination of the noise properties of the system, can
be ascribed to a finite value of $A_{11}$ and $A_{22}$.  From
Eqs. (\ref{eq:3u},\ref{eq:4u}), it is possible to write
$\left|\mathcal{G}^\theta(\omega)\right|^2$ as
\begin{align}
  \left|\mathcal{G}^\theta(\omega)\right|^2= 
                             \left|A_{11}(\omega)\right|^2 \cos \theta^2 
                             &+\left|A_{21}(\omega)\right|^2 \sin \theta^2\nonumber\\
                 &+ s_{2\theta}  \left|A_{11}A_{21}\right| s_{\Delta_1} \nonumber\\  
                 +          \left|A_{22}(\omega)\right|^2 \sin \theta^2 &+
                             \left|A_{12}(\omega)\right|^2 \cos \theta^2 \nonumber\\
                 &- s_{2\theta}  \left|A_{22}A_{12}\right| s_{\Delta_2}, 
\label{eq:10u}
\end{align}
with $s_{\Delta1}= \sin\left(\phi_{21}-\phi_{11}\right)$ and $s_{\Delta2}=
\sin\left(\phi_{12}-\phi_{22}\right)$.
Moreover, since $\left|A_{12}\right|=\left|A_{21}\right|$, we can write 
 \begin{align}
  \left|\mathcal{G}^\theta\left(\omega\right)\right|^2= 
                             \mathcal{A}_s+\mathcal{A}_x + \mathcal{A}_\Delta
   \cos \left[2 \theta + \phi \right]
\label{eq:14r}
\end{align}
with $\mathcal{A}_{s,\Delta}=\left|A_{11}\right|^2 \pm \left|A_{22}\right|^2$,
$\mathcal{A}_x=\left|A_{12}\right|^2=\left|A_{21}\right|^2$ ,
$\mathcal{A}_\phi=\sqrt{\mathcal{A}_x}\left(\left|A_{11}\right|
  s_{\Delta1}-\left|A_{12}\right|s_{\Delta2}\right)$,
$\mathcal{A}_\Delta=\sqrt{\mathcal{A}_D^2+\mathcal{A}_\phi^2}$,
$\phi=\arctan \frac{\mathcal{A}_\phi}{\mathcal{A}_D}$. Eq. \eqref{eq:14r} the
frequency-dependence of the maximum gain angle through the frequency dependence
of the added phase factor $\phi$.

\begin{figure}[ht]
\includegraphics[width=0.4\textwidth]{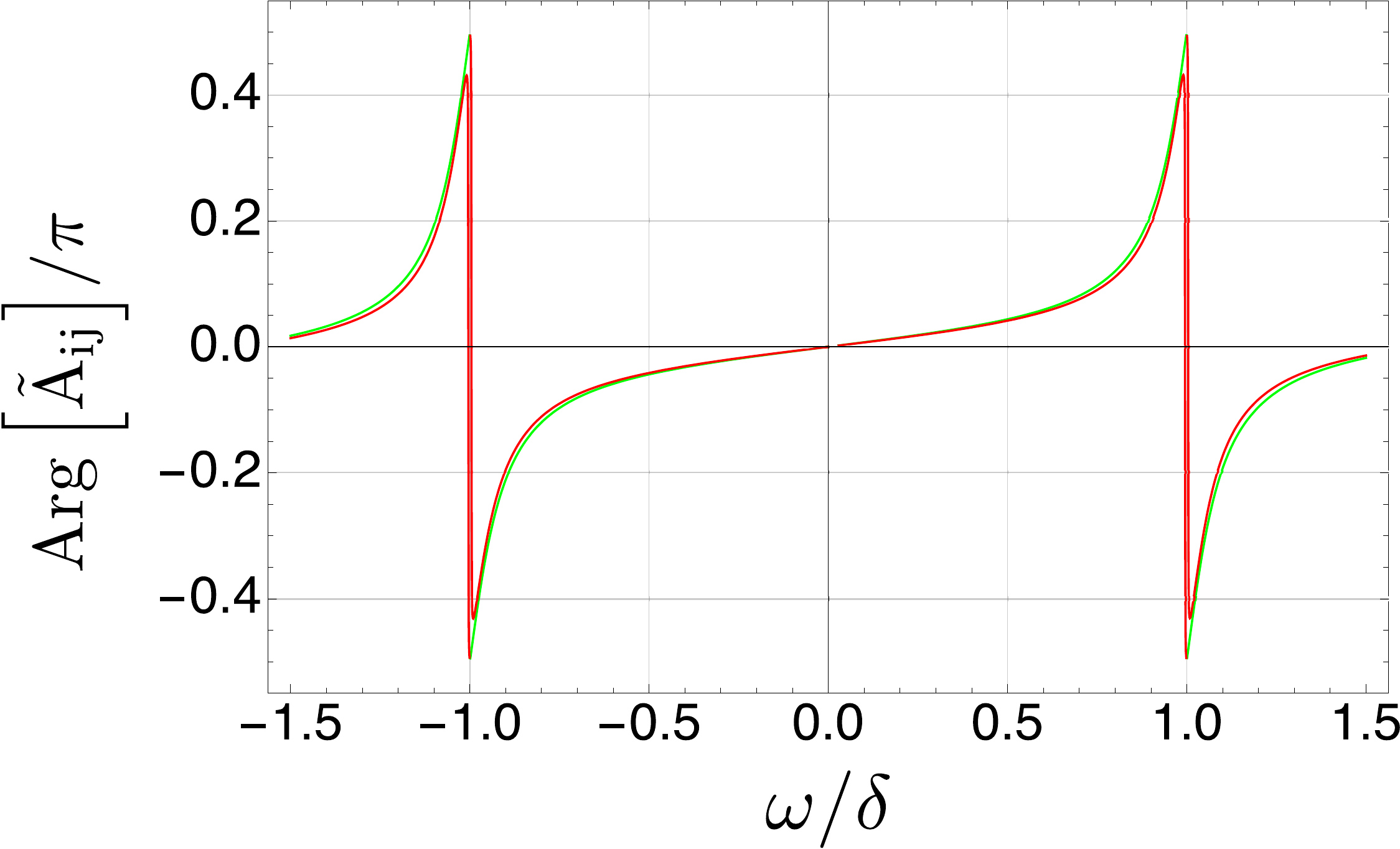}
\caption{Phase of $\tilde{A}_{11}$ (green), $\tilde{A}_{22}$ (red). For $\omega=0$
  and $\omega=\omega_{max}$ phase-sensitive amplification is possible, for other
values of $\omega$ the amplifier behaves as a phase-mixing device.}
\label{fig:phase}
\end{figure}

In order to show that away from $\omega=\omega_{max}$ the amplifier cannot be
described in terms of phase-sensitive amplification, we evaluate for the
optomechanical case the frequency dependence of the phases
$\mathrm{Arg}\left[\tilde{A}_{11}\right]$ and
$\mathrm{Arg}\left[\tilde{A}_{22}\right]$--note that
$\mathrm{Arg}\left[\tilde{A}_{12}\right]=\mathrm{Arg}\left[\tilde{A}_{21}\right]=\pm
\pi/2$, see Eq. \eqref{eq:4n}. Since for $\omega=0$,
$\tilde{A}_{21}=\tilde{A}_{12}=0$ and $\tilde{A}_{11}$, $\tilde{A}_{22}$ are
both real (see Fig. \ref{fig:phase}), the matrix $\tilde{A}$ is, in this case,
real, and therefore real-valued SVD, corresponding to a rotation to the
preferred quadratures, is possible. Analogously, for $\omega=\omega_{max}$ the
phases of all four terms are equal, implying that, in this case the matrix
$\tilde{A}$ is proportional to a real matrix and thus, again it can be rotated
to the preferred quadratures.
For all other frequency values, since all terms of matrix $\mathbf{\tilde{A}}$ are
non-vanishing and possess different phases, real-valued SVD is not possible and
thus a rotation to the preferred quadratures is not possible. 

\section{Noise properties}
\label{sec:noise-properties}

We turn now to the discussion of the added noise properties of the amplifier,
assuming that both the mechanical oscillator and the cavity field are subject to
noise --below referred to as mechanical and internal noise (see
Fig. \ref{cavity}). Otherwise stated, we assume that we can write the added
noise as $\mathcal{F}_\theta =\mathcal{F}^m_\theta + \mathcal{F}^I_\theta$,
where
\begin{align}
  \mathcal{F}_\theta^m = \left[A^m_{11} \cos \theta 
                                      - i A^m_{21} \sin \theta \right]  X_\omega^{1\,m}+ \nonumber \\
                                    \left[i A^m_{12} \cos \theta 
                                      - i A^m_{22} \sin \theta \right] X_\omega^{2\,m}
  \label{eq:19u}
\end{align}
where $A^m_{ij}$s are defined from Eq. \eqref{eq:10r}, in analogy to the
definitions given in Eq. \eqref{eq:8r} for the input signals and
% \begin{align}
%   &A^m_{11}\left(\omega\right) = \frac{1}{2}\left[\left(C_\omega+\bar{C}_\omega\right) + \left(D_\omega+\bar{D}_\omega\right)\right] \nonumber \\
%   &A^m_{21}\left(\omega\right) = \frac{1}{2}\left[\left(C_\omega-\bar{C}_\omega\right) + \left(D_\omega-\bar{D}_\omega\right)\right] \nonumber \\
%   &A^m_{12}\left(\omega\right) = \frac{1}{2}\left[\left(C_\omega-\bar{C}_\omega\right) -\left( D_\omega-\bar{D}_\omega\right)\right] \nonumber \\
%   &A^m_{22}\left(\omega\right) = \frac{1}{2}\left[\left(C_\omega+\bar{C}_\omega\right) - \left(D+\bar{D}_\omega\right)\right],
%  \label{eq:15m}
% \end{align}
$F_\theta^I$ is obtained the same way by replacing the superscript $m$ by $I$ in
Eq. \eqref{eq:19u}.
Focusing on the $\omega=\omega_{max}$ resonance, with the same approximations
as the ones used in the derivation of the gain coefficients, we have
% $D \simeq 0$ and
% $\bar{C} \simeq 0$, and  
% \begin{align}
%   C&=-\frac{2\sqrt{\gamma \kappa_e}}{\kappa} \frac{i
%      G_-}{\gamma_{eff}-i\left(\omega-\omega_{max}\right)}\nonumber \\
%   \bar{D}&=-\frac{2\sqrt{\gamma \kappa_e}}{\kappa} \frac{i G_+}{\gamma_{eff}-i\left(\omega-\omega_{max}\right)},
%   \label{eq:25}
% \end{align}
% allowing us to write \tmp{check the factors of 2}
\begin{align}
  &A^m_{11}\left(\omega\right) =A^m_{12}\left(\omega\right)
                                            =-\frac{i\sqrt{\gamma \kappa_e}}{\kappa} \frac{ G_--G_+}{\frac{\gamma_{eff}}{2}-i\left(\omega-\omega_{max}\right)}\nonumber \\
  &A^m_{22}\left(\omega\right) =A^m_{21}\left(\omega\right)
                                             =-\frac{i\sqrt{\gamma
    \kappa_e}}{\kappa} \frac{ G_-+G_+}{\frac{\gamma_{eff}}{2}-i\left(\omega-\omega_{max}\right)}.
   \label{eq:20u}
\end{align}
For the internal noise, with the same approximations considered for the
calculation of the gain, we have
\begin{align}
A^{I}_{21} &= \frac{-2\kappa_i/\kappa^2  \left(G_-+G_+\right)^2}{\frac{\gamma_{eff}}{2}- i
  \left(\omega-\omega_{max}\right)}\nonumber \\
A^{I}_{12} &=\frac{-2\kappa_i/\kappa^2 \left(G_--G_+\right)^2}{\frac{\gamma_{eff}}{2} - i
  \left(\omega-\omega_{max}\right)}\nonumber \\
A^{I}_{11}&=A^{I}_{22}=\frac{\kappa_i}{\kappa}\left[1-\frac{2 G_{BG}^2/\kappa}{\frac{\gamma_{eff}}{2} - i
  \left(\omega-\omega_{max}\right)}\right]+\frac{\kappa_i}{\kappa}. 
\label{eq:23u} 
\end{align}
With the expressions given by Eq. \eqref{eq:16r}, and excluding the possibility
of squeezed noise, we can write the contribution to the added noise as
\tmp{verify the factors of 2 p. 35 of 1808.. notes}
\begin{align}
  S_F^{\theta\,m} = 2\left[\left|A^m_{11}\right|^2 \cos \theta^2 +\left|A^m_{22}\right|^2
  \sin \theta^2\right]\left(2 n_m+1\right),
  \label{eq:21u}
\end{align}
where $n_m$ is the thermal population of the mechanical bath and analogously for
the internal cavity noise.
% where
% \begin{align}
%   &A^I_{11}\left(\omega\right) =
%     \frac{1}{2}\left[\left(A^I_\omega+\bar{A}^I_\omega\right) +
%     \left(B^I_\omega+\bar{B}^I_\omega\right)\right]=  \nonumber \\
%   &A^I_{21}\left(\omega\right) = \frac{1}{2}\left[\left(A^I_\omega-\bar{A}^I_\omega\right) + \left(B^I_\omega-\bar{B}^I_\omega\right)\right] \nonumber \\
%   &A^I_{12}\left(\omega\right) = \frac{1}{2}\left[\left(A^I_\omega-\bar{A}^I_\omega\right) -\left( B^I_\omega-\bar{B}^I_\omega\right)\right] \nonumber \\
%   &A^I_{22}\left(\omega\right) = \frac{1}{2}\left[\left(A^I_\omega+\bar{A}^I_\omega\right) - \left(B_\omega^I+\bar{B}^I_\omega\right)\right],
%  \label{eq:15I}
% \end{align}

Assuming that $\kappa_i \ll \kappa$, and $n_c^I \ll N_m$
 in the regime relevant for the experiment, corresponding to a
cavity thermal occupation of less than one quantum and to a thermal bath for the
mechanical resonator of a few hundreds quanta, the contribution from the mechanical noise
is dominant. The approximate expressions given in Eq. \eqref{eq:16r}, allow us
to write the total added noise at $\omega=\omega_{max}$ as
\begin{align}
  S_{add}^{\theta}=& \frac{S_F^{\theta\,   m}+S_F^{\theta\,I}} {\left|\mathcal{G}_\omega^\theta\right|^2} \nonumber \\ 
                   \simeq& \frac{\gamma \kappa
                                      \left[ 
                                        \left(G_-+G_+\right)^2 \sin^2 \theta +  \left(G_--G_+\right)^2 \cos^2 \theta
                                      \right]}
                    {2 \left[\left(G_-+G_+\right)^4 \sin^2 \theta+
                           \left(G_--G_+\right)^4 \cos^2 \theta\right]}  \nonumber
  \\   & \cdot\left(2n_m+1\right) 
 \label{eq:24u}
\end{align}
For $G_- \gtrsim G_+$, this expression allows establishing a condition under
which the quantum limit for phase-insensitive amplification is overcome by the
(phase-sensitive) optomechanical amplifier discussed here, namely
\begin{align}
  \left(G_-+G_+\right)^2 > \gamma \kappa (2 n_m+1) \implies S^\theta_{add}<1/2
  \label{eq:17r}
 \end{align}
 for $\theta \neq 0$.  On the other hand, even if the condition given by
 Eq. \eqref{eq:17r} is not fulfilled, it is still possible to ``beat'' the
 quantum limit in the PMA regime, reaching $S_{add,m}<1/2$ away from
 $\omega=\omega_{max}$. This relies on the different phase dependence of
 mechanical added noise and gain. Namely, the condition $A_{11}\neq 0$ allows
 for a shift in the location of the maximum of $\mathcal{G}_\omega^\theta$ as a
 function of $\theta$. Since this phase shift is absent for the added mechanical
 noise (see Fig. \ref{GainAbove}(b)), the presence of a $A_{11}\neq 0$ term
 effectively allows for a relative shift of the phases for which gain and noise
 reach their maxima.
% \begin{figure}[ht]
% \includegraphics[width=0.35\textwidth]{AddedNoiseAboveCmp.pdf}
% \caption{Total (blue), mechanical (green), internal (red) added noise for the
%   same parameters as in Fig \ref{AddedNoiseThresholdAbove3D}, and mechanical
%   linewidth $\gamma=2 10^{-5}$ for $\omega=\omega_{max}$\emph{(a)},
%   $\omega=\omega_{max}-\gamma_{eff}$.
%   \emph{(b)},$\omega=\omega_{max}+\gamma_{eff}$. \emph{(b)}}
% \label{NoiseTAbove}
% \end{figure}

Stated otherwise, it is possible to reach amplification with noise properties
below the quantum limit by shifting the input signal frequency away from
$\omega_{max}$. 
\begin{figure}[ht]
\includegraphics[width=0.5\textwidth]{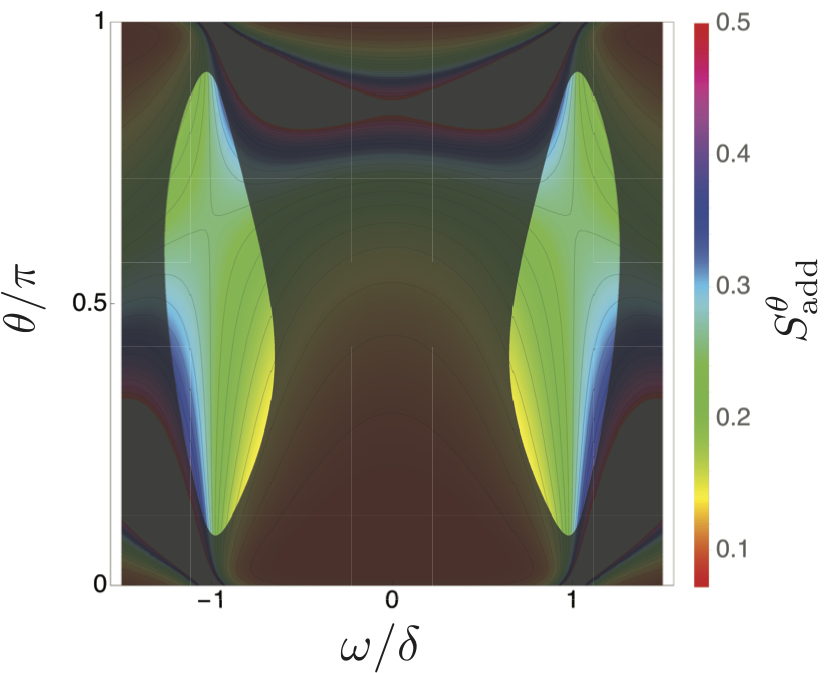}
\caption{Total added noise for a drive fulfilling the condition given in
  Eq. \eqref{eq:17r} ($n_m=200$, $n_c^I=0.1$ all other parameters as in
  Fig. \ref{GainAbove}). Grey areas correspond to regions with added noise
  larger than the AQL, as a guide to the eye, the light-colourd areas correspond
  to a  gain larger than 10.}
\label{AddedNoiseThresholdAbove3D}
\end{figure}
\begin{figure}[ht]
\includegraphics[width=0.5\textwidth]{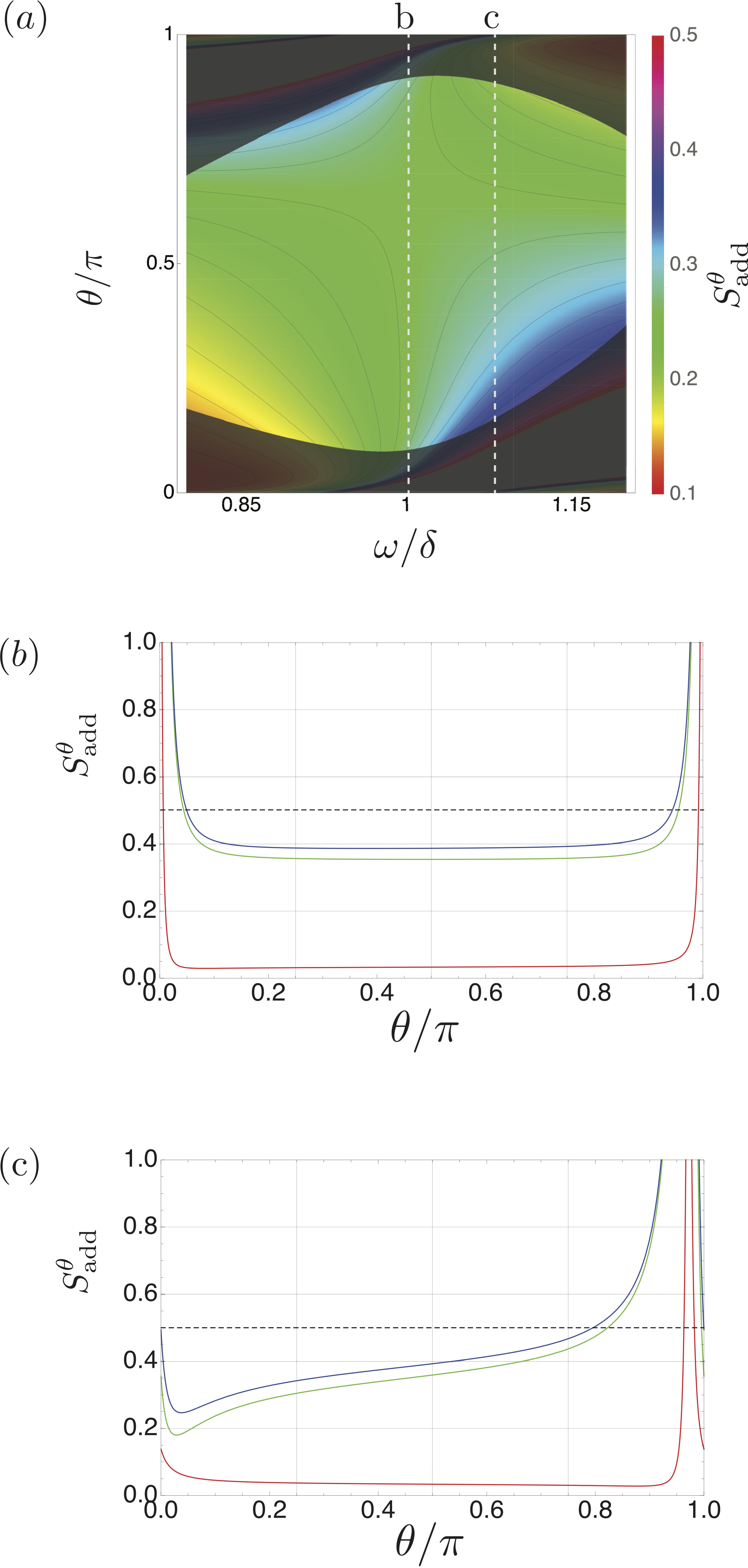}
\caption{\emph{(a)} Zoom of figure \ref{AddedNoiseThresholdAbove3D} for
  $\omega \simeq \delta$, dashed lines correspond to the plots in \emph{(b)} and
  \emph{(c)}.\emph{(b-c)} Total (blue), mechanical (green), internal (red) added
  noise for the same drive as in Fig \ref{AddedNoiseThresholdAbove3D}:
  \emph{(b)} on resonance ($\omega=\omega_{max}$),\emph{(c)} off resonance
  ($\omega=\omega_{max}+\gamma_{eff}$) .}
\label{AddedNoiseThresholdAbove}
\end{figure}
In Figs. \ref{AddedNoiseThresholdAbove3D},\ref{AddedNoiseThresholdAbove} we
depict the added noise as a function of $\omega$ and $\theta$ for a value of the
pump intensities leading to amplification with noise properties below the AQL for
$\omega=\omega_{max}$. In Fig. \ref{AddedNoiseThresholdAbove}(b),(c) it is
possible to see that shifting away from $\omega=\omega_{max}$ leads to a
reduction of the region for which  $S_{add}^{\theta}<1/2$. 
\begin{figure}[ht]
\includegraphics[width=0.5\textwidth]{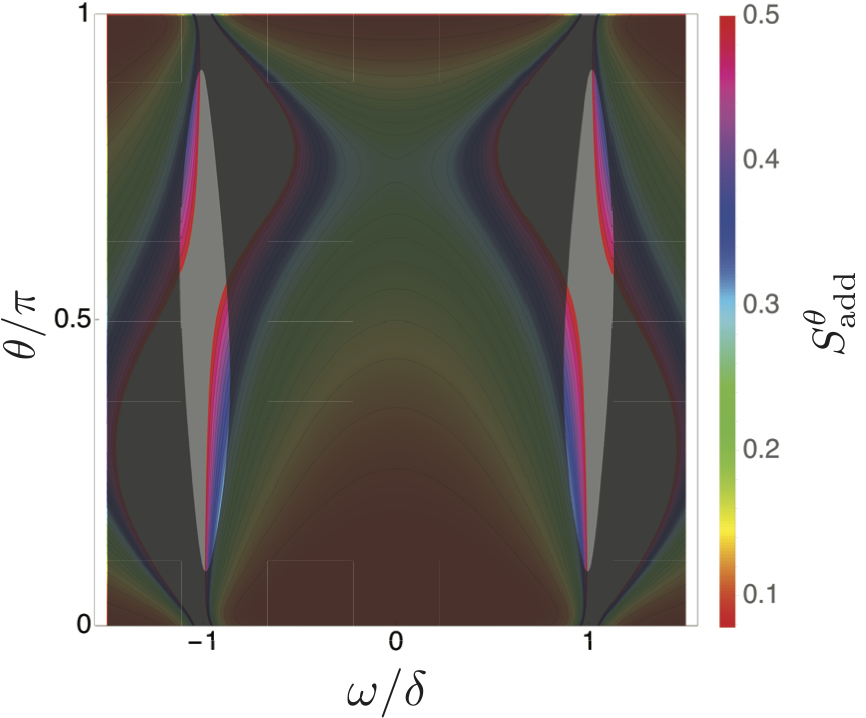}
\caption{Total added noise for a drive below $\gamma \kappa (2 n_m+1)$
  ($G_+=0.04, G_-= 0.048$, all other parameters as in the previous
  figures). Grey areas correspond to regions with added noise larger than the
  AQL}
\label{AddedNoiseThresholdBelow3D}
\end{figure}

\begin{figure}[ht]
\includegraphics[width=0.5\textwidth]{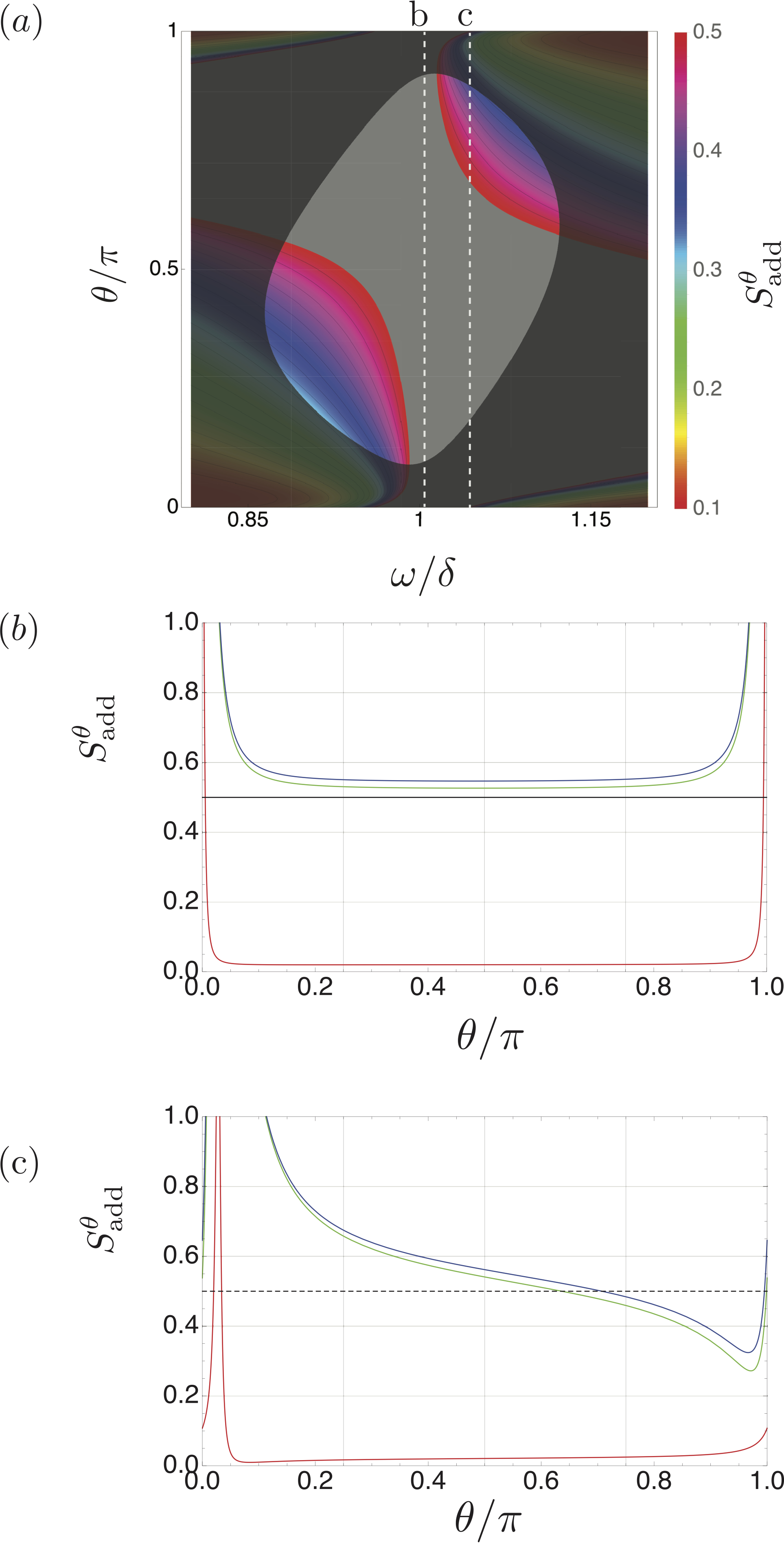}
\caption{\emph{(a)} Zoom of figure \ref{AddedNoiseThresholdBelow3D} for
  $\omega \simeq \delta$, dashed lines correspond to the plots in \emph{(b)} and
  \emph{(c)}.\emph{(b-c)} Total (blue), mechanical (green), internal (red) added
  noise for the same drive as in Fig \ref{AddedNoiseThresholdBelow3D}:
  \emph{(b)} on resonance ($\omega=\omega_{max}$),\emph{(c)} off resonance
  ($\omega=\omega_{max}+\gamma_{eff}$).}
\label{AddedNoiseThresholdBelow}
\end{figure}
In Figs. \ref{AddedNoiseThresholdBelow3D},\ref{AddedNoiseThresholdBelow}, where
we plot the total added noise for a pump leading to amplification with noise
propertiesa bove the AQL for $\omega=\omega_{max}$, the converse is true:
shifting away from $\omega=\omega_{max}$, leads to the possibility of reaching
below AQL amplification. This is a direct consequence of the different
$\theta$-dependence of the gain $\mathcal{G}^\theta(\omega)$ and the mechanical
contribution to the added noise $S^{\theta\, m}_F$.

\section{Conclusions}
\label{sec:conclusions}

We have here demonstrated a novel regime of quantum signal amplification beyond
the usual phase-insensitive/phase-sensitive amplification paradigm, which we
call phase-mixing amplification. In addition, we have provided a specific
example of phase-sensitive amplification in the context of optomechanics,
demonstrating the possibility of below-AQL amplification and showing how the
different phase dependence of gain and noise can increase the parameters range
over which below-AQL amplification is possible.

\clearpage

\appendix

\section{PMA for a coherent field}
\label{sec:appendix-iii}

In order to further clarify the concept of phase-mixing amplification, we
provide a simple example of how a phase-mixing amplifier works for an input
characterised by a coherent monochromatic signal defined around a carrier
frequency $\omega_0$ as
\begin{align}
 \braket{\mathbf{E} }  \propto x_1(t) \cos \omega_0 t + x_2(t) \sin \omega_0 t  
  \label{eq:28u}
\end{align}
where $ x_1(t)\doteq \braket{X_1}$, $ x_2(t)\doteq \braket{X_2}$ represent the
(slowly) time-varying expectation values of quadrature fields, defined with
respect to the carrier frequency $\omega_0$.  If we assume that
\begin{align}
  x_1(t)  = \Xi_1 \cos(\bar{\omega} t +\phi)  \nonumber \\
  x_2(t)  = \Xi_2 \sin(\bar{\omega} t +\phi)
  \label{eq:29u}
\end{align}
or, analogously, in frequency domain
\begin{align}
  & x_1(\omega)= \Xi_1 \left[ e^{- i \phi_1} \delta\left(\bar{\omega}-\omega\right) + e^{ i \phi_1}
  \delta\left(\bar{\omega}+\omega \right)\right] \nonumber \\
  & x_2(\omega)= \Xi_2  \left[ e^{- i \phi_2} \delta\left(\bar{\omega}-\omega\right) + e^{ i \phi_2}
    \delta\left(\bar{\omega}+\omega \right)\right], 
  \label{eq:30u}
\end{align}
(where we have set $\phi_1=\phi$ and $\phi_2=\phi-\pi/2$) and considering the I/O relations for the phase-mixing
amplifier --Eqs. \eqref{eq:9r}-- we can write the expression for the output field quadratures time dependence, defined
around the carrier frequency $\omega_0$ as for the input field, (neglecting the noise sources) as
\begin{align}
  y_{\theta}(t)=&\left\{
                           \left[
                                    A_{11}\left(\bar{\omega}\right)  \cos \theta 
                                 -i A_{21}\left(\bar{\omega}\right)  \sin \theta
                           \right] e^{-i\left(\bar{\omega}t+\phi_1\right)}
                  +\right.
                     \nonumber \\
                     & \left. \left[
                                    A_{11}\left(-\bar{\omega}\right)  \cos \theta 
                                 -i A_{21}\left(-\bar{\omega}\right)  \sin \theta
                           \right] e^{i\left(\bar{\omega}t+\phi_1\right)}
                   \right\} \Xi_1 + \nonumber \\
                   & \left\{
                           \left[
                                  i  A_{12}\left(\bar{\omega}\right)  \cos \theta 
                                 + A_{22}\left(\bar{\omega}\right)  \sin \theta
                           \right] e^{-i\left(\bar{\omega}t+\phi_2\right)}+
                     \right. \nonumber \\
                   &\left.  \left[
                                  i A_{12}\left(-\bar{\omega}\right)  \cos \theta 
                                 + A_{22}\left(-\bar{\omega}\right)  \sin \theta
                           \right] e^{i\left(\bar{\omega}t+\phi_2\right)}
                   \right\} \Xi_2.
  \label{eq:31u}
\end{align}
Since $A_{ij}(\bar{\omega})=A^*_{ij}(-\bar{\omega})$, Eq. \eqref{eq:31u} can be
written as 
\begin{align}
   y_{\theta\,t}= & \left[  \left|A_{11}(\omega)\right| \cos \theta \cos(\bar{\omega}t +
                    \bar{\phi}_{11}) \right. \nonumber \\
                        &\left.  -  \left|A_{21}(\omega)\right| \sin \theta  \sin(\bar{\omega}t + \bar{\phi}_{21})
                      \right] \Xi_1 + \nonumber \\
                      & \left[   \left|A_{22}(\omega)\right| \sin \theta \cos(\bar{\omega}t +
                        \bar{\phi}_{22}) \right. \nonumber \\
                      &\left.    + \left|A_{12}(\omega)\right| \cos \theta  \sin(\bar{\omega}t + \bar{\phi}_{12})
                      \right] \Xi_2,
  \label{eq:26r}
\end{align}
where we have defined $\bar{\phi}_{\rm ij} = \phi_{\rm j}-\phi_{\rm ij}$.
Eq. \eqref{eq:26r} can be written also as
\begin{align}
  y_{\theta\,t}= & \left[ \mathcal{A}_1 \cos\left(\bar{\omega}t + \bar{\phi}_1^\theta\right)\, \Xi_1 +
                                \mathcal{A}_2 \sin\left(\bar{\omega}t + \bar{\phi}_2^\theta\right)\, \Xi_2  
                          \right]
 \label{eq:33u}
\end{align}
having defined
\begin{align}
  \mathcal{A}_1&=\sqrt{|A_{11} \cos \theta|^2+|A_{21} \sin \theta|^2+ A_{11}A_{21}
                 s_{2 \theta} \sin\left(\bar{\phi}_{11}-\bar{\phi}_{21}\right)} \nonumber \\
  \mathcal{A}_2&=\sqrt{|A_{22} \sin \theta|^2+|A_{12} \cos \theta|^2- A_{12}A_{22}
                 s_{2 \theta} \sin\left(\bar{\phi}_{22}-\bar{\phi_{12}}\right)}
  \label{eq:34u}
\end{align}
and
\begin{align}
  \bar{\phi}_1^\theta = \arctan 
    \frac{A_{11} \cos \theta \sin \bar{\phi}_{11} + A_{21} \sin \theta  \cos \bar{\phi}_{21} }
           {A_{11} \cos \theta \cos \bar{\phi}_{11} - A_{21}  \sin \theta \sin \bar{\phi}_{21} }
            \nonumber \\
  \bar{\phi}_2^\theta =\arctan 
     \frac {A_{12} \cos \theta \sin \bar{\phi}_{12}+A_{22} \sin \theta  \cos \bar{\phi}_{22} }
             {A_{12} \cos \theta \cos \bar{\phi}_{12}-A_{22}  \sin \theta \sin \bar{\phi}_{22} }. 
  \label{eq:24r}
\end{align}
Finally, from Eq. \eqref{eq:24r}, it is possible to write 
\begin{align}
   y_{\theta\,t}= \mathcal{A}_\theta  \cos(\bar{\omega}t+\eta_\theta) \Xi  
  \label{eq:25r}
\end{align}
with 
\begin{align}
  \mathcal{A}_\theta \Xi=\sqrt{\mathcal{A}_1^2\, \Xi_1^2+\mathcal{A}_2^2 \,\Xi_2^2+
  2\mathcal{A}_1\mathcal{A}_2 \Xi_1 \Xi_2\sin \left(\bar{\phi}_2-\bar{\phi}_1\right)} 
  \label{eq:35u}
\end{align}
and
\begin{align}
  \eta_\theta =\arctan\left[
                              \frac{\mathcal{A}_1 \sin \bar{\phi}_1 - \mathcal{A}_2 \cos \bar{\phi}_2}
                                     {\mathcal{A}_1 \cos \bar{\phi}_1 - \mathcal{A}_2 \sin \bar{\phi}_2}  
                             \right].
  \label{eq:26br}
\end{align}

In the case of real coefficients $A_{\rm ij}$, Eq. \eqref{eq:26br} allows to evaluate the expression for the
output quadratures $y_1\doteq y_{\theta=0}, y_2\doteq y_{\theta=\pi/2}$ as
\begin{align}
  y_{1\,t}=\sqrt{A_{11}^2  \Xi_1^2 + A_{12}^2 \Xi_2^2 } \cos(\bar{\omega}t +\phi)   \nonumber \\
  y_{2\,t}=\sqrt{A_{22}^2  \Xi_2^2 + A_{21}^2 \Xi_1^2 }  \sin\left(\bar{\omega}t + \phi \right)
  \label{eq:37u}
\end{align}
clearly showing how each output quadrature depends on the amplitude of both
input quadratures.

The analysis performed above represents a simple example of how each
output quadrature depends on both input quadratures, allowing thus to tailor the
phase properties of the output signal with far reaching consequences, as we
outline in the analysis of the specific optomechanical device proposed here.

\section{Derivation of the I/O EOMs}
\label{sec:appendix-i}

The mechanical degrees of freedom can be eliminated from Eq. \eqref{eq:4r},
leading to the following equation for the Bogoliubov mode $\alpha$
\begin{align}
   -i \omega \alpha_\omega = G_{{BG}}^2 \chi_m
  \alpha_\omega &-\frac{\kappa}{2} \alpha_\omega \nonumber \\ 
  &   + \sqrt{\kappa} \alpha_{{in}\, \omega} - i G_{BG} \chi_m  \sqrt{\gamma} b_{{in}\,\omega}
 \label{eq:25u}
\end{align}
which can be solved to give 
\begin{align}
  \label{eq:26u}
   \alpha_\omega =\chi_c^{\mathrm{eff}} \sqrt{\tilde{\kappa}}\tilde{\alpha}_{in\,\omega}. 
\end{align}
where
\begin{align}
   \sqrt{\tilde{\kappa}} \tilde{\alpha}_{in\,\omega}=\sqrt{\kappa} \alpha_{in\,\omega}-i G_{BG}
\chi_m \sqrt{\gamma}b_{in\,\omega}.
  \label{eq:7}
\end{align}

%\bibliographystyle{apsrev_abb}
%\bibliography{NewAmpli.bib}

\begin{thebibliography}{27}
\expandafter\ifx\csname natexlab\endcsname\relax\def\natexlab#1{#1}\fi
\expandafter\ifx\csname bibnamefont\endcsname\relax
  \def\bibnamefont#1{#1}\fi
\expandafter\ifx\csname bibfnamefont\endcsname\relax
  \def\bibfnamefont#1{#1}\fi
\expandafter\ifx\csname citenamefont\endcsname\relax
  \def\citenamefont#1{#1}\fi
\expandafter\ifx\csname url\endcsname\relax
  \def\url#1{\texttt{#1}}\fi
\expandafter\ifx\csname urlprefix\endcsname\relax\def\urlprefix{URL }\fi
\providecommand{\bibinfo}[2]{#2}
\providecommand{\eprint}[2][]{\url{#2}}

\bibitem[{\citenamefont{Clerk et~al.}(2010)\citenamefont{Clerk, Devoret,
  Girvin, Marquardt, and Schoelkopf}}]{Clerk:2010dh}
\bibinfo{author}{\bibfnamefont{A.~A.} \bibnamefont{Clerk}},
  \bibnamefont{et~al.}, \bibinfo{journal}{Rev. Mod. Phys.}
  \textbf{\bibinfo{volume}{82}}, \bibinfo{pages}{1155} (\bibinfo{year}{2010}).

\bibitem[{\citenamefont{Zurek}(1991)}]{ZUREK:1991td}
\bibinfo{author}{\bibfnamefont{W.~H.} \bibnamefont{Zurek}},
  \bibinfo{journal}{Phys Today} \textbf{\bibinfo{volume}{44}},
  \bibinfo{pages}{36} (\bibinfo{year}{1991}).

\bibitem[{\citenamefont{Abbott et~al.}(2016)\citenamefont{Abbott, Abbott,
  Abbott, Abernathy, Acernese, Ackley, Adams, Adams, Addesso, Adhikari
  et~al.}}]{Abbott:2016ki}
\bibinfo{author}{\bibfnamefont{B.~P.} \bibnamefont{Abbott}},
  \bibnamefont{et~al.}, \bibinfo{journal}{Phys. Rev. Lett.}
  \textbf{\bibinfo{volume}{116}}, \bibinfo{pages}{061102}
  (\bibinfo{year}{2016}).

\bibitem[{\citenamefont{Haus and Mullen}(1962)}]{Anonymous:rI5t-iRf}
\bibinfo{author}{\bibfnamefont{H.~A.} \bibnamefont{Haus}} \bibnamefont{and}
  \bibinfo{author}{\bibfnamefont{J.}~\bibnamefont{Mullen}},
  \bibinfo{journal}{Phys. Rev.} \textbf{\bibinfo{volume}{128}},
  \bibinfo{pages}{2407} (\bibinfo{year}{1962}).

\bibitem[{\citenamefont{Caves}(1982)}]{Caves:1982wq}
\bibinfo{author}{\bibfnamefont{C.~M.} \bibnamefont{Caves}},
  \bibinfo{journal}{Phys. Rev. D} \textbf{\bibinfo{volume}{26}},
  \bibinfo{pages}{1817} (\bibinfo{year}{1982}).

\bibitem[{\citenamefont{Wootters and Zurek}(1982)}]{Wootters:1982ex}
\bibinfo{author}{\bibfnamefont{W.~K.} \bibnamefont{Wootters}} \bibnamefont{and}
  \bibinfo{author}{\bibfnamefont{W.~H.} \bibnamefont{Zurek}},
  \bibinfo{journal}{Nature} \textbf{\bibinfo{volume}{299}},
  \bibinfo{pages}{802} (\bibinfo{year}{1982}).

\bibitem[{\citenamefont{Castellanos-Beltran
  et~al.}(2008)\citenamefont{Castellanos-Beltran, Irwin, Hilton, Vale, and
  Lehnert}}]{CastellanosBeltran:2008cg}
\bibinfo{author}{\bibfnamefont{M.~A.} \bibnamefont{Castellanos-Beltran}},
  \bibnamefont{et~al.}, \bibinfo{journal}{Nat. Phys.}
  \textbf{\bibinfo{volume}{4}}, \bibinfo{pages}{929} (\bibinfo{year}{2008}).

\bibitem[{\citenamefont{Bergeal et~al.}(2010)\citenamefont{Bergeal, Schackert,
  Metcalfe, Vijay, Manucharyan, Frunzio, Prober, Schoelkopf, Girvin, and
  Devoret}}]{Bergeal:2010iu}
\bibinfo{author}{\bibfnamefont{N.}~\bibnamefont{Bergeal}},
  \bibnamefont{et~al.}, \bibinfo{journal}{Nature}
  \textbf{\bibinfo{volume}{465}}, \bibinfo{pages}{64} (\bibinfo{year}{2010}).

\bibitem[{\citenamefont{Zhong et~al.}(2013)\citenamefont{Zhong, Menzel,
  Di~Candia, Eder, Ihmig, Baust, Haeberlein, Hoffmann, Inomata, Yamamoto
  et~al.}}]{Zhong:2013ca}
\bibinfo{author}{\bibfnamefont{L.}~\bibnamefont{Zhong}}, \bibnamefont{et~al.},
  \bibinfo{journal}{New J Phys} \textbf{\bibinfo{volume}{15}},
  \bibinfo{pages}{125013} (\bibinfo{year}{2013}).

\bibitem[{\citenamefont{Massel et~al.}(2011)\citenamefont{Massel, Heikkil{\"a},
  Pirkkalainen, Cho, Saloniemi, Hakonen, and
  Sillanp{\"a}{\"a}}}]{Massel:2011ca}
\bibinfo{author}{\bibfnamefont{F.}~\bibnamefont{Massel}}, \bibnamefont{et~al.},
  \bibinfo{journal}{Nature} \textbf{\bibinfo{volume}{480}},
  \bibinfo{pages}{351} (\bibinfo{year}{2011}).

\bibitem[{\citenamefont{Metelmann and Clerk}(2014)}]{Metelmann:2014bp}
\bibinfo{author}{\bibfnamefont{A.}~\bibnamefont{Metelmann}} \bibnamefont{and}
  \bibinfo{author}{\bibfnamefont{A.~A.} \bibnamefont{Clerk}},
  \bibinfo{journal}{Phys. Rev. Lett.} \textbf{\bibinfo{volume}{112}},
  \bibinfo{pages}{133904} (\bibinfo{year}{2014}).

\bibitem[{\citenamefont{Metelmann and Clerk}(2015)}]{Metelmann:2015gb}
\bibinfo{author}{\bibfnamefont{A.}~\bibnamefont{Metelmann}} \bibnamefont{and}
  \bibinfo{author}{\bibfnamefont{A.~A.} \bibnamefont{Clerk}},
  \bibinfo{journal}{Phys. Rev. X} \textbf{\bibinfo{volume}{5}},
  \bibinfo{pages}{021025} (\bibinfo{year}{2015}).

\bibitem[{\citenamefont{T{\'o}th et~al.}(2016)\citenamefont{T{\'o}th, Bernier,
  Nunnenkamp, Glushkov, Feofanov, and Kippenberg}}]{Toth:2016vk}
\bibinfo{author}{\bibfnamefont{L.~D.} \bibnamefont{T{\'o}th}},
  \bibnamefont{et~al.}, \bibinfo{journal}{arXiv:}\eprint{1602.05180}, (\bibinfo{year}{2016}).

\bibitem[{\citenamefont{Ockeloen-Korppi
  et~al.}(2016)\citenamefont{Ockeloen-Korppi, Damsk{\"a}gg, Pirkkalainen,
  Heikkil{\"a}, Massel, and Sillanp{\"a}{\"a}}}]{OckeloenKorppi:2016uy}
\bibinfo{author}{\bibfnamefont{C.~F.} \bibnamefont{Ockeloen-Korppi}},
  \bibnamefont{et~al.}, \bibinfo{journal}{arXiv:}\eprint{1602.05779}
  (\bibinfo{year}{2016}), to appear in PRX.

\bibitem[{\citenamefont{Ralph et~al.}(2009)\citenamefont{Ralph, Lund, and
  Lvovsky}}]{Ralph:bx}
\bibinfo{author}{\bibfnamefont{T.~C.} \bibnamefont{Ralph}},
  \bibinfo{author}{\bibfnamefont{A.~P.} \bibnamefont{Lund}}, \bibnamefont{and}
  \bibinfo{author}{\bibfnamefont{A.}~\bibnamefont{Lvovsky}}, in
  \emph{\bibinfo{booktitle}{QCMC: Ninth International Conference on QCMC}}
  (\bibinfo{publisher}{AIP}, \bibinfo{year}{2009}), pp.
  \bibinfo{pages}{155--160}.

\bibitem[{\citenamefont{L{\"a}hteenm{\"a}ki
  et~al.}(2014)\citenamefont{L{\"a}hteenm{\"a}ki, Vesterinen, Hassel, Paraoanu,
  Sepp{\"a}, and Hakonen}}]{Lahteenmaki:2014gw}
\bibinfo{author}{\bibfnamefont{P.}~\bibnamefont{L{\"a}hteenm{\"a}ki}},
  \bibnamefont{et~al.}, \bibinfo{journal}{J. Low Temp. Phys.}
  \textbf{\bibinfo{volume}{175}}, \bibinfo{pages}{868} (\bibinfo{year}{2014}).

\bibitem[{\citenamefont{Gardiner and Collett}(1985)}]{Gardiner:1985ig}
\bibinfo{author}{\bibfnamefont{C.}~\bibnamefont{Gardiner}} \bibnamefont{and}
  \bibinfo{author}{\bibfnamefont{M.~J.} \bibnamefont{Collett}},
  \bibinfo{journal}{Phys. Rev. A} \textbf{\bibinfo{volume}{31}},
  \bibinfo{pages}{3761} (\bibinfo{year}{1985}).

\bibitem[{Note1()}]{Note1}
Note1, \bibinfo{note}{the subscript $\omega $ stands for the frequency, and we
  use the Fourier convention where $a^\dagger _\omega $ is the conjugate of
  $a_\omega $.}

\bibitem[{\citenamefont{Walls and Milburn}(2008)}]{Walls:2008em}
\bibinfo{author}{\bibfnamefont{D.~F.} \bibnamefont{Walls}} \bibnamefont{and}
  \bibinfo{author}{\bibfnamefont{G.~J.} \bibnamefont{Milburn}},
  \emph{\bibinfo{title}{{Quantum optics}}} (\bibinfo{publisher}{Springer Berlin
  Heidelberg}, \bibinfo{year}{2008}).

\bibitem[{\citenamefont{Aspelmeyer et~al.}(2014)\citenamefont{Aspelmeyer,
  Kippenberg, and Marquardt}}]{Aspelmeyer:2014ce}
\bibinfo{author}{\bibfnamefont{M.}~\bibnamefont{Aspelmeyer}},
  \bibinfo{author}{\bibfnamefont{T.~J.} \bibnamefont{Kippenberg}},
  \bibnamefont{and}
  \bibinfo{author}{\bibfnamefont{F.}~\bibnamefont{Marquardt}},
  \bibinfo{journal}{Rev. Mod. Phys.} \textbf{\bibinfo{volume}{86}},
  \bibinfo{pages}{1391} (\bibinfo{year}{2014}).

\bibitem[{\citenamefont{Caves et~al.}(1980)\citenamefont{Caves, Thorne, Drever,
  Sandberg, and Zimmermann}}]{Caves:1980jpa}
\bibinfo{author}{\bibfnamefont{C.~M.} \bibnamefont{Caves}},
  \bibnamefont{et~al.}, \bibinfo{journal}{Rev. Mod. Phys.}
  \textbf{\bibinfo{volume}{52}}, \bibinfo{pages}{341} (\bibinfo{year}{1980}).

\bibitem[{\citenamefont{Braginsky et~al.}(1980)\citenamefont{Braginsky,
  Vorontsov, and Thorne}}]{Braginsky:1980hj}
\bibinfo{author}{\bibfnamefont{V.~B.} \bibnamefont{Braginsky}},
  \bibinfo{author}{\bibfnamefont{Y.~I.} \bibnamefont{Vorontsov}},
  \bibnamefont{and} \bibinfo{author}{\bibfnamefont{K.~S.}
  \bibnamefont{Thorne}}, \bibinfo{journal}{Science}
  \textbf{\bibinfo{volume}{209}}, \bibinfo{pages}{547} (\bibinfo{year}{1980}).

\bibitem[{\citenamefont{Clerk et~al.}(2008)\citenamefont{Clerk, Marquardt, and
  Jacobs}}]{Clerk:2008je}
\bibinfo{author}{\bibfnamefont{A.~A.} \bibnamefont{Clerk}},
  \bibinfo{author}{\bibfnamefont{F.}~\bibnamefont{Marquardt}},
  \bibnamefont{and} \bibinfo{author}{\bibfnamefont{K.}~\bibnamefont{Jacobs}},
  \bibinfo{journal}{New J Phys} \textbf{\bibinfo{volume}{10}},
  \bibinfo{pages}{095010} (\bibinfo{year}{2008}).

\bibitem[{\citenamefont{Hertzberg et~al.}(2009)\citenamefont{Hertzberg,
  Rocheleau, Ndukum, Savva, Clerk, and Schwab}}]{Hertzberg:hf}
\bibinfo{author}{\bibfnamefont{J.~B.} \bibnamefont{Hertzberg}},
  \bibnamefont{et~al.}, \bibinfo{journal}{Nat. Phys.}
  \textbf{\bibinfo{volume}{6}}, \bibinfo{pages}{213} (\bibinfo{year}{2009}).

\bibitem[{\citenamefont{Wollman et~al.}(2015)\citenamefont{Wollman, Lei,
  Weinstein, Suh, Kronwald, Marquardt, Clerk, and Schwab}}]{Wollman:2015gx}
\bibinfo{author}{\bibfnamefont{E.~E.} \bibnamefont{Wollman}},
  \bibnamefont{et~al.}, \bibinfo{journal}{Science}
  \textbf{\bibinfo{volume}{349}}, \bibinfo{pages}{952} (\bibinfo{year}{2015}).

\bibitem[{\citenamefont{Pirkkalainen et~al.}(2015)\citenamefont{Pirkkalainen,
  Damsk{\"a}gg, Brandt, Massel, and Sillanp{\"a}{\"a}}}]{Pirkkalainen:2015ki}
\bibinfo{author}{\bibfnamefont{J.~M.} \bibnamefont{Pirkkalainen}},
  \bibnamefont{et~al.}, \bibinfo{journal}{Phys. Rev. Lett.}
  \textbf{\bibinfo{volume}{115}}, \bibinfo{pages}{243601}
  (\bibinfo{year}{2015}).

\bibitem[{\citenamefont{Lecocq et~al.}(2015)\citenamefont{Lecocq, Clark,
  Simmonds, Aumentado, and Teufel}}]{Lecocq:2015dk}
\bibinfo{author}{\bibfnamefont{F.}~\bibnamefont{Lecocq}}, \bibnamefont{et~al.},
  \bibinfo{journal}{Phys. Rev. X} \textbf{\bibinfo{volume}{5}},
  \bibinfo{pages}{041037} (\bibinfo{year}{2015}).

\end{thebibliography}

 \end{document}